\documentclass[pra,preprint,showpacs]{revtex4}
\usepackage[centertags]{amsmath}
\usepackage[utf8]{inputenc}
\usepackage{amsfonts}
\usepackage{amssymb}
\usepackage{amsthm}
\usepackage{amsmath}
\usepackage{newlfont}
\usepackage{epsfig}
\usepackage{amscd}
\usepackage{graphicx}
\usepackage{epsfig}
\usepackage{footnote}
\usepackage{lipsum}
\usepackage{color}
\usepackage{xcolor}
\usepackage{graphicx}
\usepackage{multirow}
\usepackage{enumerate}
\usepackage{caption}
\usepackage{subcaption}

\usepackage{amscd}

\newcommand{\beq}{\begin{equation}}
\newcommand{\eeq}{\end{equation}}
\newcommand{\ba}{\begin{array}}
\newcommand{\ea}{\end{array}}
\newcommand{\bea}{\begin{eqnarray}}
\newcommand{\eea}{\end{eqnarray}}
\begin{document}

\begin{center}
{\large \sc \bf {Optimal {{remote}} restoring of quantum states in communication lines via local magnetic field}
}

\vskip 15pt

{\large
E.B.~Fel'dman$^{1,2}$,  A.N.~Pechen$^{2,3}$ and  A.I.~Zenchuk$^{1,2}$
}

\vskip 8pt

{\it $^1$
Federal Research Center of Problems of Chemical Physics and Medicinal Chemistry RAS,
Chernogolovka, Moscow reg., 142432, Russia}.

{\it $^2$ Department of Mathematical Methods for Quantum Technologies, Steklov Mathematical Institute of Russian Academy of Sciences, Gubkina str. 8, Moscow 119991, Russia}

{\it $^3$National University of Science and Technology MISIS, Leninski prosp. 4, Moscow 119991, Russia}

{\it $^*$Corresponding author. E-mail:  zenchuk@itp.ac.ru}
\vskip 8pt

\end{center}
\date{\today}

\begin{abstract}
Optimal state transport across spin chains, which are proposed as quantum wires for information transfer in solid state quantum architectures, is an important topic for quantum technologies. In this work, we study {the remote restoring of a quantum state transferred along a spin chain.} The structural state-restoring technique provides proportionality between the appropriate elements of the density matrices of the initial sender state and receiver state at some time instant. We develop a {remote} state-restoring  protocol which uses an inhomogeneous  magnetic field  
with step-wise time-dependent Larmor frequencies as the state-control tool. For simulating the multiparametric Hamiltonian we use two approximating models. First model is based on the Trotter-Suzuki  method, while the second model is based on using short pulses of high intensity.  In both cases we estimate the accuracy of the approximation and find the optimal restoring parameters (Larmor frequencies) of the protocol which maximize the coefficients in the proportionality for spin chains of various lengths.
\end{abstract}

{\bf Keywords:} state-restoring protocol, XX-Hamiltonian, time-dependent Larmor frequencies,  Trotterization method, high-intensity short pulses

\maketitle

\section{Introduction}
\label{Section:introduxtion}

Quantum information is an intensively developing direction in quantum physics, and information transmission along spin chains, which were proposed as quantum wires for information transfer in solid state quantum architectures, is a particularly important topic for modern quantum technologies~\cite{Schleich2016,Acin2018}. We can select three different processes  associated with information propagation along spin chain. (i) Teleportation of arbitrary state \cite{BBCJPW,BPMEWZ,BBMHP}; (ii) arbitrary state transfer \cite{Bose,CDEL,KS,GKMT,GMT,ZASO,ZASO2,ZenchukJPA2012}; (iii) remote state creation~\cite{PBGWK,PBGWK2,DLMRKBPVZBW,PSB,LH,Z_2014,BZ_2015}. Optimization of information transport along spin chains is an important direction in quantum control necessary for quantum technologies~\cite{Koch2022}. {Many results on state transfer along spin-chain quantum wires have been obtained, including high fidelity communication of quantum states through one dimensional rings of qubits
with fixed interactions~\cite{Osborne2004}, fast high-fidelity information transmission using either a single local on-off switch actuator~\cite{SchirmerPRA2009} or controls applied to the two ends of the lattice~\cite{Ashhab2015}, communication at the quantum speed limit~\cite{MurphyPRA2010,Ashhab2012,OsendaPLA2021}, preparation of generic many-body states~\cite{MorigiPRL2015}, realizing a functional donor chain~\cite{Mohiyaddin2016}, free control of information transmission in chaotically kicked spin chains~\cite{AubourgJPB2016}, conditional controlled state transfer in Heisenberg XXZ spin chains via periodic drives~\cite{ShanSciRep2018}, non-adiabatic cutting and stitching of a spin chain via local control~\cite{PyshkinNJP2018}, propagation of excitations with designed site-dependent interaction strengths~\cite{FerronPS2022}, experiments with realization of universal control and error correction for multi-qubit spin registers in nitrogen-vacancy center~\cite{Taminiau2014},  Floquet prethermalization in dipolar spin chains~\cite{Peng2021}, coherent control of a nuclear spin via interactions with a rare-Earth ion~\cite{Uysal2023}, etc.} Many detailed aspects about controlling spin systems can be found in the book~\cite{KuprovBook}.

{ 
Our paper  concerns the problem of
 remote
restoring of the 
state transferred from} sender to receiver along a spin chain {and develops the recent ideas of  Refs.~\cite{FZ_2017,BFZ_Arch2018,Z_2018,FPZ_2021,BFLP_2022}.} 
{We consider spin chain such that its sender part is 
initially prepared in (arbitrary) state $\rho^{(S)}(0)$ and all other spins are 
in the ground states. Then the spin chain evolves { in time $t$} under the action of a 
magnetic field with some time-dependent Hamiltonian up to some time instant 
$t_{\mathrm{reg}}$ which we shall call time instant for state registration. 
We want that at this time instant elements of the density matrix of the receiver 
part of the spin chain $\rho^{(R)}(t_{\mathrm{reg}})$ should be proportional 
to the appropriate elements of the sender's initial state $\rho^{(S)}(0)$:}
\begin{eqnarray}\label{restoring}
\rho^{(R)}_{ij}(t_{\mathrm{reg}})=\lambda_{ij}\rho^{(S)}_{ij}(0).
\end{eqnarray}
{ Here the parameters $\lambda_{ij}$  (we also call them 
$\lambda$-parameters) do not depend on the initial state $\rho^{(S)}(0)$ 
and can be different for different $i,j$. In addition, 
Eq.~(\ref{restoring}) must be consistent with the trace-normalization of 
the density matrix, which causes certain constraints on the structure of 
the sender's initial state $\rho^{(S)}(0)$~\cite{FPZ_2021,BFLP_2022}, 
unless we smooth the requirements~(\ref{restoring}) for some (at 
least one) of the diagonal elements of the density matrix. So, equality~
(\ref{restoring}) should be satisfied either for some restricted set of 
initial states $\rho^{(S)}(0)$ and all elements, or for all initial 
states and all elements except for some (at least one) diagonal element(s). Then our goal is to find maximal in absolute value parameters $\lambda_{ij}$ which satisfy all the above properties including Eq.~(\ref{restoring}). For this, it is sufficient to consider maximization of the minimal among all $\lambda$-parameters. The $\lambda$-parameters depend on $t_{\mathrm{reg}}$, and the corresponding time instant $t_{\mathrm{reg}}$ for which this maximum is achieved is the optimal time instant for state registration.}

{The protocol of remote restoring the transferred state}  relies on special unitary transformation applied to the so-called extended receiver which includes the nodes of the receiver and several neighboring nodes. The dimensionality of the extended receiver predicts the number of the free parameters in the unitary transformation which can be effectively used to reach the goal of state restoring. 
{ Thus, our remote state restoring protocol combines the ideas of state transfer (ii) and remote state creation (iii), utilizing the method of controlling the receiver's state via unitary transformations of the extended receiver \cite{Z_2014,BZ_2015}.}

However, generating this special unitary transformation is a non-trivial problem by itself. According to the Solovay-Kitaev theorem \cite{NCh}, any unitary transformation can be generated using a fixed number of one- and two-qubit operations. However, in general, this process is not effective because the number of elementary transformations needed to simulate the required unitary transformation increases exponentially (in general) with the size of this transformation. Therefore, finding methods for a direct generation of the required special unitary transformation is of high practical significance. 

In our paper we modify the state restoring protocol by { replacing the unitary transformation of the extended receiver with} the time-dependent inhomogeneous magnetic field  in the Hamiltonian. We require that this magnetic field acts selectively on some nodes of the spin chain and thus performs the goal of state restoring allowing to avoid introduction of the special unitary  transformation at the receiver side. Thus, the local intensities of this magnetic field play the role of the parameters of the unitary transformation restoring the transferred state. In this case, the restoring transformation is incorporated into the evolution operator and can not be selected as a separate unitary transformation, unlike in~\cite{BFLP_2022}. {We derive general formulas which can be applied to various Hamiltonians including general XXZ models and apply them to the particular case which is XX model.}

We consider the state-transfer communication line as a linear open homogeneous spin chain consisting of sender, transmission line and receiver which is embedded into the extended receiver.  The whole communication line is governed by the $XX$-Hamiltonian with non-zero time-dependent  Larmour frequencies at the nodes of the extended receiver (all other Larmour frequencies are zero). As a simplification of the time-dependent Hamiltonian, we adopt the piece-wise constant Larmour frequencies similar to that used for single qubit and two qubits~\cite{PP_2023,MP_2023},  thus splitting  the required time interval  into  the set of subintervals where the Hamiltonian is constant, so that formally we can construct exact evolution operator over each of these intervals.

{ Thus, the main subject of our manuscript is constructing the remote state-restoring protocol with piece-constant Larmour frequencies used as a controlling tool.}
However, the protocol requires diagonalization of the Hamiltonian parameterized by free parameters, { which is a large matrix with symbolic matrix elements. It can be diagonalized analytically up to $4\times 4$ dimension. In our work, we consider matrices wth dimensions up to $30\times 30$ for which direct symbolic calculation of the evolution operator is hardly possible.} To overcome this difficulty, we propose two models. 
\begin{itemize}
\item[---]
{\bf Model 1} {{uses as controls piecewise constant functions} and is based on the 
Trotter-Suzuki approach~\cite{Trotter,Suzuki} for computation of the evolution separating the diagonal time-dependent part of the Hamiltonian 
from the constant flip-flop part. This is a theoretical approximation of the actual Hamiltonian evolution with step-wise Larmour frequencies. Physically it corresponds to switching among different constant values of magnetic field.}
\item[---]
{\bf Model 2} {uses as controls almost impulse functions, that allows to split free and control Hamiltonians for computation of the evolution operator. Physically it corresponds to using sequences of short high-intensity  pulses of the local magnetic field which leads to large Larmour frequencies localized in time. In this case Larmour frequencies prevail over the rest of the Hamiltonian  so that the latter can be neglected over the short intervals of the magnetic field pulses. From experimental point of view the privilege of this method is the possibility of its realization by the tools of multi-quantum NMR \cite{Baum}}.
\end{itemize}
{ We note that the above models can not be reduced  to each other. Typical controls for the Models 1 and 2 are shown on Fig.~\ref{Fig:step}.  Emphasize that using step-wise and impulse controls in these models has been made for technical simplicity in calculating the evolution operator. However, more general forms of controls can be used for the protocol as well.

\begin{figure*}[!]
\centering
\includegraphics[scale=0.65]{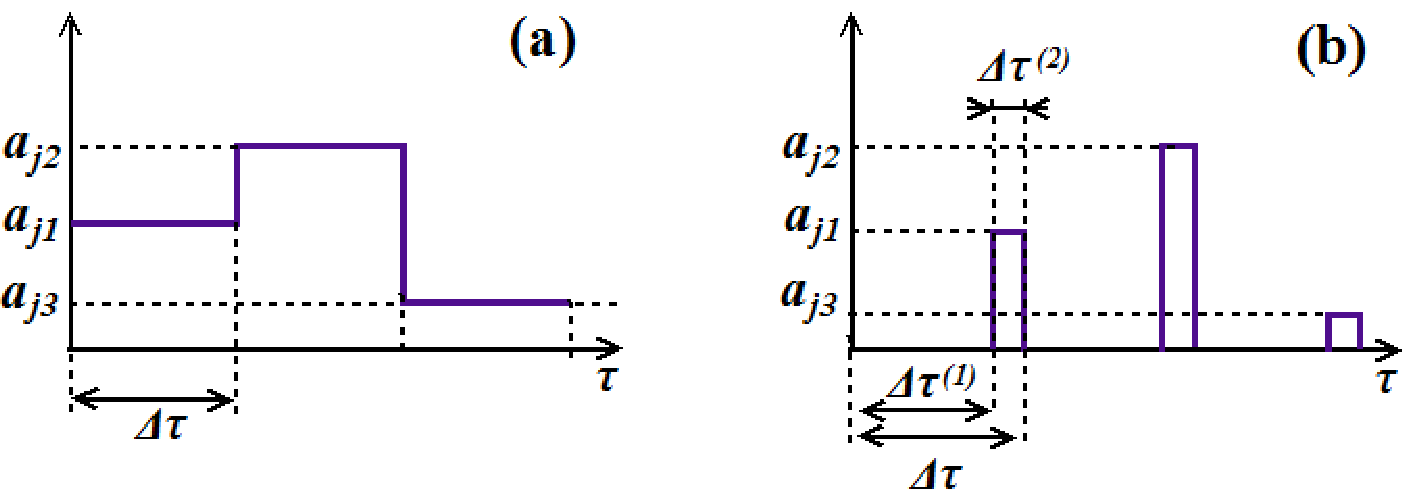}
\caption{{ Typical examples of time evolution of the controls $\omega_j$ ($\tau$ is some rescaled time defined below): (a) for Model 1, where $\omega_j$ is constant over each interval $\Delta\tau$; (b) for Model 2, where each interval $\Delta\tau$ is divided in long subinterval $\Delta\tau^{(1)}$ with zero control magnetic field and short subinterval  $\Delta\tau^{(2)}$ with sufficiently strong magnetic field.}}  
\label{Fig:step}
\end{figure*}

For these two simulation models, one could consider two separate questions. First is to apply any of the models to the analysis of the control protocol. Second is to compare the two models in the context of quantum dynamics without relation to the control protocol. These are two unrelated questions, and our work considers the first one.}

An often used method for optimization of controls for spin systems is GRadient Ascent Pulse Engineering (GRAPE) approach. {Originally developed for finding NMR pulse sequences for quantum systems}~\cite{Khaneja2005}, it was later extended and applied to various systems~\cite{GlaserJPB2011,Fouquieres2011,PechenTannor2011,Lucarelli2018,PP_2023}. Recently, GRAPE approach was applied to the analysis of quantum control landscapes for single spins with Hamiltonian dynamics and driven by only coherent control in~\cite{VMP_21} and for single spins interacting with the environment and driven by both coherent control and the environmental drive in~\cite{PP_2023}. Open GRAPE was developed to realize a CNOT with high fidelities~\cite{GlaserJPB2011}. However, for large spin systems dealing with large matrices might be computationally quite expensive. In this regard, a perspective of using GRAPE approach to this state restoring protocol is an important prospect for a future work. 

The paper is organized as follows. In Sec.~\ref{Section:T}, we consider the general state-restoring protocol based on the time-dependent Larmor frequencies at the nodes of the extended receiver. Evolution under the time-dependent XX-Hamiltonian and two model approximating this evolution are discussed in Sec.~\ref{Section:ev}. State restoring of (0,1)-excitation states with numerical simulation is given in Sec.~\ref{Section:examples}. Concluding remarks are presented in Sec.~\ref{Section:conclusions}.

\section{State restoring evolution}
\label{Section:T}

Governed by some Hamiltonian $H$ evolution of a density matrix $\rho$ is described by the Liouville-von Neumann equation (where we set Planck constant $\hbar=1$):
\begin{eqnarray}\label{rho}
i \rho_t(t) = [H(t),\rho(t)]=H(t)\rho(t) - \rho(t) H(t).
\end{eqnarray}
If $H(t)=H_0$ is independent on $t$, then this equation can be simply integrated:
\begin{eqnarray}
\rho(t)=e^{-iH_0 t} \rho(0) e^{iH_0 t},
\end{eqnarray}
where $\rho(0)$ is the initial state. Otherwise, the analytical solution is given by the time-ordered (chronological)  exponent and generically can be explicitly solved using numerical methods. 

We consider a one-dimensional communication line including the sender $S$ {($N^{(S)}$ nodes)}, receiver $R$ {($N^{(R)}=N^{(S)}$ nodes)} and transmission line $TL$  {($N^{(TL)}$ nodes)} connecting them. 
We aim on solving the initial value problem with the
initial state
\begin{eqnarray}\label{rho0}
\rho(0)=\rho^{(S)}(0)\otimes \rho^{(TL;R)}(0),
\end{eqnarray}
where $\rho^S(0)$ is an arbitrary  initial sender's state to be transferred to the receiver $R$, while the initial state of the transmission line and receiver $\rho^{(TL;R)}(0)$ is the ground state,
\begin{eqnarray}
\rho^{(TL;R)}(0) = |0_{TL,R}\rangle
{\langle 0_{TL,R}|} = {\mbox{diag}}(1,0,\dots).
\end{eqnarray}

Then, formally we can represent the evolution $\rho$ as follows:
\begin{eqnarray}\label{rhonm}
\rho(t)=U(t)\rho(0) U^\dagger(t),
\end{eqnarray}
where $U(t)$ is the unitary ($U(t) U^\dagger(t) = E$, where $E$ is the identity matrix) evolution operator which is the solution of the Schr\"odinger equation 
\begin{eqnarray}\label{Krho}
i \frac{d U(t)}{dt} = H(t) U(t)
\end{eqnarray}
with the initial condition 
\begin{eqnarray}\label{ic}
U(0)=E.
\end{eqnarray}

We  use the Hamiltonian describing spins in the strong $z$-directed magnetic field with the time-dependent diagonal part
\begin{eqnarray}\label{HH0}
H(t)&=&H_0 +\Omega(t),\quad \Omega(t)= \sum_{j=0}^N \omega_j(t) Z_{j},\\\label{L}
&&
\omega_j =\gamma B_j  ,\quad{ Z_i=\frac{1}{2}\left(\begin{array}{cc}
1&0\cr0&-1\end{array}
\right)},
\end{eqnarray}
where $H_0$ is the constant part of the Hamiltonian (free Hamiltonian), {$Z_{i}$} is the $z$-projection of the $i$th spin momentum, $w_j$ are Larmor frequencies, $\gamma $ is the gyromagnetic ratio, $B_j$ is the  local magnetic field at the $j$th spin.
We also require 
\begin{eqnarray}\label{com0}
[H,Z]=0,\quad Z= \sum_i Z_{i},
\end{eqnarray}
i.e, the evolution should preserve the $z$-projection of the total spin momentum.
In turn, commutation relation (\ref{com0}) assumes the block-diagonal structure of the Hamiltonian
\begin{eqnarray}\label{bH}
H={\mbox{diag}} (H^{(0)},H^{(1)},\dots ),
\end{eqnarray}
where the  block $H^{(j)}$ governs the evolution of the $j$-excitation subspace and $H^{(0)}$ is a scalar. 
Without loss of generality, we set 
\begin{eqnarray}\label{bH0}
H^{(0)}=0,
\end{eqnarray}
i.e., 0-excitation state of the spin chain does not evolve under Hamiltonian~(\ref{HH0}).
We consider the restoring problem for the  nondiagonal elements of the receiver density matrix $\rho^{(R)}$, 
i.e. creation of such receiver matrix $\rho^{(R)}$ that
\begin{eqnarray}\label{rest}
\rho^{(R)}_{ij}(t_{\mathrm{reg}}) = \lambda_{ij} \rho^{(S)}_{ij}(0), \quad i\neq j,
\end{eqnarray}
where the receiver density matrix $\rho^{(R)}$ is obtained by taking partial trace:
\begin{eqnarray}\label{R}
\rho^{(R)}={\mbox{Tr}}_{S,TL}(\rho).
\end{eqnarray}
We shall emphasize that the scale parameters $\lambda_{ij}$ are universal, i.e., they do not depend on the elements of  the initial sender's density matrix and are completely defined by the Hamiltonian and the selected time instant for state registration {$t_{\mathrm{reg}}$}.  

Let us write condition (\ref{rest}) in terms of the operator $U$. 
To this end, it is convenient to introduce the multi-index notation 
\begin{eqnarray}
I=\{I_S I_{TL} I_R\},
\end{eqnarray}
where each subscript indicates the appropriate subsystem: sender $S$, transmission line $TL$ and receiver $R$. Each multi-index $I_S$, $I_{TL}$ and $I_R$ is the ordered set of binary digits whose number equals to the number of spins in the appropriate subsystem, each digit takes value either 1 (excited spin) or 0 (spin in the ground state). By $|I|$, $|I_S|$ and etc. we denote the cardinality of the corresponding set of indices, {i.e.} number of units in the corresponding multi-index.
Then Eq.~(\ref{R}) can be written as follows:
\begin{eqnarray}\label{rhoR}
\rho^{(R)}_{N_R;M_R} = \sum_{I_S,J_S} T_{N_RM_R;I_SJ_S}\rho^{(S)}_{I_S;J_S} ,
\end{eqnarray}
where matrix elements of $T$ are expressed in terms of matrix elements of $U$ as follows:
\begin{eqnarray}\label{TT}
T_{N_RM_R;I_SJ_S}&\equiv&\sum_{{{N_S,N_{TL}}\atop{|N_S|+|N_{TL}|=}}\atop{|I_S|-|N_R|=|J_S|-|M_R|}} U_{N_SN_{TL}N_R; I_S0_{TL}0_R} U^\dagger_{J_S 0_{TL}0_R; N_SN_{TL}M_R}.
\end{eqnarray}
{The tensor $T$ transfers the initial  density matrix of the sender $S$  to the receiver $R$  and is represented by a composition of Kraus operators  $K_{N_{TL}N_R}$ \cite{KrausBook} defined by the elements $U_{N_SN_{TL}N_R; I_S0_{TL}0_R}$. Notice, that there are $N^{(TL)} N^{(R)}$ of such operators in (\ref{TT}), and only $(N^{(R)})^2$ of them are independent (usually $N^{(TL)}>N^{(R)}$), but we do not need to select independent operators in our paper.}

{ Writing the sum in (\ref{TT})}, we take into account the constraint $|N|=|I|$ for subscripts in  $U_{N;I}$ that  is required by  commutation relation (\ref{com0}). Next, we have to satisfy the restoring constraint (\ref{rest}) for the non-diagonal elements of $\rho^{(R)}$ implying  assumption (\ref{bH0}), which yields
\begin{eqnarray}
U_{0_S0_{TL}0_R; 0_S0_{TL}0_R}=1.
\end{eqnarray}
Restoring constraint  (\ref{rest})  presumes  the following definitions of  the $\lambda$-parameters:
\begin{eqnarray}\label{K}
\lambda_{N_R0_R} &=&T_{N_R0_R;N_R0_R}= T_{0_RN_R;0_RN_R}^* =U_{0_S0_{TL} N_R; N_R;0_{TL} 0_R}  ,\\\nonumber
\lambda_{N_RM_R}&=&T_{N_RM_R;N_RM_R} =
U_{0_S0_{TL}N_R; N_R0_{TL}0_R} U^\dagger_{M_R 0_{TL}0_R; 0_S0_{TL}M_R}=\\\nonumber
&&\lambda_{N_R0_R}\lambda_{0_RM_R} =  
 \lambda_{N_R0_R}\lambda^*_{M_R0_R}
\end{eqnarray}
and  the following expressions for the other elements of $T$:
\begin{eqnarray}\label{constr1}
T_{N_RM_R;N_RJ_S} &=&U_{0_S0_{TL}N_R; N_R0_{TL}0_R} U^\dagger_{J_S 0_{TL}0_R; 0_S0_{TL}M_R}=\\\nonumber
&&\lambda_{N_R0_R} U^\dagger_{J_S 0_{TL}0_R; 0_S0_{TL}M_R},\;\;J_S\neq M_R\\\label{constr2}
T_{N_RM_R;I_SM_R} &=&U_{0_S0_{TL}N_R; I_S0_{TL}0_R} U^\dagger_{M_R 0_{TL}0_R; 0_S0_{TL} M_R} = \\\nonumber
&&\lambda^*_{M_R0_R} U_{0_S0_{TL}N_R; I_S0_{TL}0_R}, \;\; I_S\neq N_R,\\\label{constr3}
T_{N_RM_R;I_SJ_S} &=& \sum_{{{N_S,N_{TL}}\atop{|N_S|+|N_{TL}|=}}\atop{|I_S|-|N_R|=|J_S|-M_R|}} U_{N_SN_{TL}N_R; I_S0_{TL}0_R} U^\dagger_{J_S 0_{TL}0_R; N_SN_{TL}M_R},\\\nonumber
&&N_R \neq I_S,\;\; M_R\neq J_S.
\end{eqnarray}
Formulae (\ref{K})-(\ref{constr3})  hold for any $N_R$ and $M_R$. But for $N_R\neq M_R$ (nondiagonal elements of $\rho^{(R)}$) at time instant of state registration $t_{\mathrm{reg}}$ we have the constraints
\begin{eqnarray}\label{constr0}
T_{N_RM_R;N_RJ_S} = T_{N_RM_R;I_SM_R} =T_{N_RM_R;I_SJ_S}=0,\;\;N_R \neq I_S,\;\; M_R\neq J_S ,
\end{eqnarray}
which  are equivalent to
\begin{eqnarray}\label{P1}
 &&U_{0_S0_{TL}N_R; I_S0_{TL}0_R} =0,\quad N_R\neq I_S; \quad \\\nonumber
&&T_{N_RM_R;I_SJ_S}\equiv\sum_{{{N_S,N_{TL}}\atop{|N_S|+|N_{TL}|=}}\atop{|I_S|-|N_R|=|J_S|-M_R|}} U_{N_SN_{TL}N_R; I_S0_{TL}0_R} U^\dagger_{J_S 0_{TL}0_R; N_SN_{TL}M_R} =0,\\\nonumber
&& N_R\neq I_S,\quad M_R\neq J_S,\quad N_R\neq M_R.
\end{eqnarray}
Notice that the first constraint in~(\ref{P1}) prescribes a diagonal form to the matrix
$\hat U$ with the elements $\hat U_{N_R; I_S} = U_{0_S0_{TL}N_R; I_S0_{TL}0_R}$.

To satisfy constraints (\ref{P1}), we introduce  the set of  $N^{(ER)}$ ($ER$ means the Extended Receiver) nonzero controls 
$\omega_i(t)$, $i=N-N^{(ER)}+1, \dots, N$,
assuming
\begin{eqnarray}
\omega_i=0,\quad  i=1,\dots,N-N^{(ER)}.
\end{eqnarray}
Non-zero $\omega_i(t)$ are some functions of $t$. Let them be step-functions~\cite{PP_2023}
\begin{eqnarray}\label{LF}
\omega_k(t) = \sum_{j=1}^{K_\omega} a_{kj} \theta_j(t),\quad  \theta_j(t) =\left\{\begin{array}{ll}
1, &t_{j-1} < t\le t_j\cr
0  & {\mbox{otherwise}}. \end{array}
\right.
\end{eqnarray}
Here we split the entire time interval 
$[0,t_{\mathrm{reg}}]$ 
in $K_\omega$ intervals of different (in general) lengths. 
As a simplest variant, let us fix $t_j$ and consider  $a_{kj}$ as control parameters. Then the evolution operator can be split into the set of non commuting (in general) operators $U_j$ of form
\begin{eqnarray}\label{VV}
U_j(t) &=& \left\{ 
\begin{array}{ll}
\exp(-i H_j t),& t_{j-1} < t\le t_j\cr
1  & {\mbox{otherwise}}
\end{array}
\right.,\\\label{Hj}
&&
H_j = H_0 +\Omega_j,\\\label{Omega}
&&
\Omega_j=\sum_i a_{ij} I_{zi},\;\;j=1,\dots,K_\omega.
\end{eqnarray} 
Thus, $U = U_{K_\omega}\dots U_1$ and
\begin{eqnarray}\label{UtK}
U(t_{\mathrm{reg}})=  U_{K_\omega}(\Delta t_j)\dots U_1(\Delta t_1) U(0),\quad  \Delta t_j = t_j -t_{j-1}.
\end{eqnarray}
The number of control parameters equals $N^{(ER)} K_\omega$. Hereafter we identify {$t_{K_\omega}\equiv t_{\mathrm{reg}}$} for simplicity. 

\section{Evolution under XX-Hamiltonian}
\label{Section:ev}
We consider the evolution under the XX-Hamiltonian with inhomogeneous $t$-dependent magnetic field, i.e., in~(\ref{HH0}) free Hamiltonian is
\begin{eqnarray}\label{XX}
H_0&=&\sum_{j>i} D_{ij} (X_{i}X_{j}+Y_{i}Y_{j}  ),\;\; {
X_i=\frac{1}{2}\left(\begin{array}{cc}
0&1\cr
1&0
\end{array}\right),\;\;  Y_i=\frac{1}{2}\left(\begin{array}{cc}
0&-i\cr
i&0
\end{array}\right)} ,  \\\label{com}
&&[H_0,I_z]=0.
\end{eqnarray}
{Here $D_{ij}=\gamma^2/r_{ij}^3$ are the coupling constants between the $i$th and $j$th spins (for $\hbar=1$), $\gamma$ is the gyromagnetic ratio, 
$r_{ij}$  is the distance between the $i$th and $j$th spins}, and {$X_,\;\;Y_i$ are  the operators of, respectively, the $x$- and $y$-projections of the $i$th spin.} 

{ Notice that Hamiltonian (\ref{XX}) includes interaction among all nodes of the chain. 
The spin chain model involving XX and YY terms with nearest neighbour interactions and generally different (anisotropic) couplings was investigated in~\cite{Lieb1961} where it was called XY model since it involves 'x' and 'y' components of spin operators. More general XYZ anisotropic form with long range coupling as above  was derived in~\cite{Porras2004}. For isotropic XX and YY couplings and anisotropic ZZ coupling such model is also called as XXZ~\cite{Maik2012}. In other references, for instance in~\cite{PM,IBRR}, the Hamiltonian with nearest neighbor interaction and equal coefficients ahead of $XX$ and $YY$ terms is called $XX$-Hamiltonian. It is a particular case of the XXZ model of~\cite{Porras2004,Maik2012} without ZZ term. Here we adopt the term XX-model to the Hamiltonian with all node interactions and isotropic $XX$ and $YY$ couplings.}
Due to the commutation relation~(\ref{com}), the Hamiltonian $H$ has the block-diagonal form (\ref{bH})
with
\begin{eqnarray}\label{H0}
H^{(0)} = \sum_{i=N-N^{(ER)}+1}^N \omega_i.
\end{eqnarray}
{Since there is a freedom in the energy of the ground state (that can be controlled by the external homogeneous magnetic field $B$ adding the term $\Omega Z$ to the Hamiltonian, $Z=\sum_i Z_{i}$, $\Omega=\gamma B$), we can} require $H^{(0)}=0$ that yields the constraint on control functions $\omega_i$,
\begin{eqnarray}\label{omega2}
\sum_{i=N-N^{(ER)}+1}^N \omega_i =0,
\end{eqnarray}
and the corresponding constraints on control parameters $a_{kj}$:
\begin{eqnarray}\label{a}
\sum_{k=N-N^{(ER)}+1}^{N} a_{kj}=0,\;\;1\le j\le K_\omega.
\end{eqnarray}
We set
\begin{eqnarray}\label{a2}
a_{Nj}=-\sum_{k=N-N^{(ER)}+1}^{N-1} a_{kj},\;\;1\le j\le K_\omega.
\end{eqnarray}
Therefore, the extended receiver with $N^{(ER)}$ nodes has $K_\omega(N^{(ER)}-1)$ free parameters. 
Although  formally  implementing  the evolution with  piece-constant Larmor  frequencies (\ref{LF}) is very simple, 
its simulation faces the problem of  Hamiltonian diagonalization because of the free parameters $a_{kj}$ which can not be fixed until solving system (\ref{P1}) {and therefore must be treated symbolically}. Therefore we are forced to turn to the approximate evolution which would simplify the dependence on the free parameters. 
Below we consider two models allowing to include the Larmour frequencies in a simpler way. We use dimensionless time $\tau = D_{12} t$ and equal intervals 
\begin{eqnarray}\label{dtau}
\Delta\tau = \tau_j-\tau_{j-1} ={ \frac{\tau_{\mathrm{reg}}}{K_\omega}.}
\end{eqnarray}

\subsection{Model 1: Trotterization} 
\label{Section:Trot}
As the first model, we propose the Troterrization method for simulating the evolution using the Trotter-Suzuki approach~\cite{Trotter,Suzuki}. 
{For this model we considet the $\tau$-dependence of the Larmour frequencies $w_j$ in the form of the  step-functions (\ref{LF}), 
see Fig.\ref{Fig:step}a},
and  approximate  the evolution operator as follows:
\begin{eqnarray}\label{Vtr}
U_j(\Delta\tau) = e^{-i \frac{H_j}{D_{12}} \Delta \tau}\approx U^{(n)}_j (\Delta\tau) =
\left(e^{-i \frac{{H_0}}{D_{12}}\frac{\Delta\tau}{n}}
e^{-i \frac{\Omega_j}{D_{12}}\frac{\Delta\tau}{n}}\right)^n, 
\end{eqnarray}
where $n$ is the Trotterization number. Thus, we replace the operator  $U$  by $U^{(n)}$ in (\ref{UtK}),
\begin{eqnarray}\label{Ktr2}
U^{(n)}(\tau_{\mathrm{reg}})=\prod_{j=1}^{K_\omega} \left(U_j\Big(\frac{\Delta\tau}{n}\Big)\right)^{n}=
\prod_{j=1}^{K_\omega} \left(U_j\Big(\frac{\tau_{\mathrm{reg}}}{K_\omega n}\Big)\right)^{n},
\end{eqnarray}
and introduce the superscript $^{(n)}$  in Eqs.~(\ref{K}) and (\ref{P1}):
\begin{eqnarray}\label{lambda2}
&& \lambda^{(n)}_{N_RM_R}=  
 \lambda^{(n)}_{N_R0_R}(\lambda^{(n)}_{M_R0_R})^*,\;\;N_R\neq M_R,
\\\nonumber
&& \lambda^{(n)}_{N_R0_R} = U^{(n)}_{0_S0_{TL} N_R; N_R0_{TL} 0_R},
\end{eqnarray}
\begin{eqnarray}\label{constr12}
&&
U^{(n)}_{0_S 0_{TL}N_R; I_S0_{TL}0_R}(\omega,\tau_{\mathrm{reg}})  =0,\;\; N_R\neq I_S,\\\nonumber
&&\sum_{N_S,N_{TL}} U^{(n)}_{N_SN_{TL}N_R; I_S0_{TL}0_R}(\omega,\tau_{\mathrm{reg}}) (U^{(n)}(\omega,\tau_{\mathrm{reg}}))^\dagger_{J_S 0_{TL}0_R; N_SN_{TL}M_R} =0,\\\nonumber
&&N_R \neq I_S,\;\; M_R\neq J_S,\;\; N_R\neq M_R,
\end{eqnarray}
Let $\omega= \{a_{kj}\}$ be a set of free parameters produced by Larmor frequencies in the operator $U^{(n)}$. System (\ref{constr12}) can be viewed as a system for the parameters $\omega$. We denote any parameters $\omega$  solving  (\ref{constr12}) by $\omega^{(n)}$ ($n$ is the Trotterization number).
However,  solution $\omega^{(n)}$ of  system~(\ref{constr12}) is not unique. 
Let $\omega^{(n,m)}$, $m=1,2,\dots$,  be the set of different solutions. 
Each  $\omega^{(n,m)}$ approximates solution $\omega$ of system 
(\ref{P1}). 

Thus, approximation by Trotterization method incorporates the Larmor frequencies in a simpler way allowing to simulate dynamics in longer chains. Now we have to estimate the accuracy of such approximation by substituting $\omega^{(n,m)}$ into (\ref{P1}). In the case of perfect approximation we would obtain identities (all left hand sides would be zero). Therefore, the following function { of $\tau_{\mathrm{reg}}$} can characterize the accuracy of Trotterization: 
\begin{eqnarray}\label{epsilon}
S^{(n)}_1(\tau) =\min_m \max_P P(\omega^{(n;m)},\tau_{\mathrm{reg}}),
\end{eqnarray}
where
\begin{eqnarray}\label{P2}
 P(\omega,\tau_{\mathrm{reg}})&:=&\{\left|U_{0_S0_{TL}N_R; I_S0_{TL}0_R}(\omega,\tau_{\mathrm{reg}})\right|,\;\;N_R\neq I_S; \;\;\\\nonumber
&&
\left|\sum_{{{N_S,N_{TL}}\atop{|N_S|+|N_{TL}|=}}\atop{|I_S|-|N_R|=|J_S|-M_R|}} U_{N_SN_{TL}N_R; I_S0_{TL}0_R} (\omega,\tau)U^\dagger_{J_S 0_{TL}0_R; N_SN_{TL}M_R}(\omega,\tau)\right|,\\\nonumber
&& N_R\neq I_S,\;\;M_R\neq J_S,\;\;N_R\neq M_R\}.
\end{eqnarray}
We emphasize that formula (\ref{P2}) involves $U$ rather then $U^{(n)}$.

Another function is associated with the accuracy of constructing the $\lambda$-parameters, i.e.,
\begin{eqnarray}\label{epsilon2}
S^{(n)}_2(\tau_{\mathrm{reg}}) =\min_m\max_{N_R} \left|\lambda^{(n)}_{N_R0_R}(\omega^{(n,m)},\tau_{\mathrm{reg}})) -
\lambda_{N_R0_R}(\omega^{(n,m)},\tau_{\mathrm{reg}}))\right|,
\end{eqnarray}
which is also zero in the case of perfect approximation.
The following two characteristics associated with $S_1^{(n)}$ and $S_2^{(n)}$ can be also useful:
\begin{eqnarray}\label{S3}
&&
S^{(n)}_3(\tau_{\mathrm{reg}}) = \max_{0\le\tilde \tau\le \tau_{\mathrm{reg}}} S^{(n)}_1(\tilde \tau),
\\\label{S4}
&&
S^{(n)}_4(\tau_{\mathrm{reg}}) = \max_{0\le\tilde \tau\le \tau_{\mathrm{reg}}} S^{(n)}_2(\tilde \tau).  
\end{eqnarray}

{ We emphasize that the functions $S^{(n)}_1(\tau)$ and $S^{(n)}_2(\tau)$ indicate the local (in time) accuracies of solving system (\ref{constr1}) and calculating $\lambda$-factors. 
On the contrary, $S^{(n)}_3(\tau)$ and $S^{(n)}_4(\tau)$ are the
global characteristics of  a model informing  what are the best accuracies  $S^{(n)}_1$ and $S^{(n)}_2$ over the interval $[0,\tau]$ at fixed chain length $N$ and sender/receiver dimension $N^{(S)}$.}
{We also} notice that the functions $S^{(n)}_i$, $i=1,\dots,4$, characterize the accuracy of Trotterization  and therefore they were not used in~\cite{BFLP_2022}. Now we introduce the function 
$S^{(n)}_5(\tau_{\mathrm{reg}})$,
\begin{eqnarray}\label{S5}
S^{(n)}_5(\tau_{\mathrm{reg}}) = \max_{0\le\tilde \tau\le \tau_{\mathrm{reg}}} \max_{m} \min_{N_R} \left|\lambda ^{(n)}_{N_R0_R} (\omega^{(n;m)},\tilde \tau)\right|,
\end{eqnarray}
which shows how accurately we have restored the transferred state. Similar function was used in \cite{BFLP_2022}. Of course, $S^{(n)}_5$ depends on the chain length $N$, which is another argument of this  function, $S^{(n)}_5(\tau,N)$. The function $S^{(n)}_5(T(N),N)$ with some fixed large enough time-interval $T$ depending on $N$ characterizes the effectiveness of application of state-restoring protocol to the chains of different lengths over the  time interval $T$.  With $S^{(n)}_5(T,N)$, we associate the time instant { for  state registration}  ${\tau_{\mathrm{reg}} =} \,\tau_0(N)$,
$0<\tau_0(N)< T $, such that 
\begin{eqnarray}\label{lamn2}
\lambda^{(n)}=S^{(n)}_5(T,N) = S^{(n)}_5(\tau_0(N) ,N).
\end{eqnarray}
In other words, $\tau_0$ is the time instance inside of the above interval $T$ at which the state restoring provides the best optimization result.

\subsection{Model 2: XX-Hamiltonian with strong magnetic pulses}
\label{Section:Model2}
We also propose another model simulating the state restoring using the inhomogeneous magnetic field. 
Similar to the Model~1 in Sec.~\ref{Section:Trot}, we split the interval $\tau$ into the set of $K_\omega$ subintervals $\Delta \tau$. But now, in addition,  we split 
each $\Delta \tau$ into 2 subintervals,  
\begin{eqnarray}\label{subint}
\Delta \tau= \Delta \tau^{(1)}+ \Delta \tau^{(2)}, 
\end{eqnarray}
so that 
\begin{eqnarray}\label{HHj}
H_j=\left\{ \begin{array}{ll}
H_0, & \tau \in \Delta\tau^{(1)}\cr
H_0 + \Omega_j, & \tau \in \Delta\tau^{(2)}
\end{array}
\right.
\end{eqnarray}
{Thus, for this model,  we consider the $\tau$-dependence of the Larmour frequencies $\omega_j$  in the pulse form shown in Fig.\ref{Fig:step}b.}
We require 
\begin{eqnarray}\label{avar}
&&
\min_{i,j} a_{ij} \gg \max_{i,j} D_{ij}, \\\nonumber
&&\max_{i, j}a_{ij} \Delta \tau^{(2)}\sim \max_{ij}D_{ij} \Delta \tau^{(1)},
\end{eqnarray}
so that 
$$
\frac{\|\Omega_j\|}{D_{12}} \Delta\tau^{(2)} \sim \frac{\|H_0\|}{D_{12}} \Delta\tau^{(1)}  ,
$$
where $\|\cdot\|$ is some matrix norm. In other words,  large $\|\Omega_j\|$s act over the short time interval. This means that the term $\Omega_j$  in the sum $H_j=H_0+\Omega_j$ (see Eq.~(\ref{VV})) dominates over $H_0$ on the interval $\Delta\tau^{(2)}$. Therefore we can approximate Hamiltonian~(\ref{HHj}) as
\begin{eqnarray}\label{HHj2}
H_j\approx \left\{ \begin{array}{ll}
H_0, & \tau \in \Delta\tau^{(1)}\cr
\Omega_j, & \tau \in \Delta\tau^{(2)}
\end{array}
\right.
\end{eqnarray}
and write the approximate evolution operator $U_j(\Delta \tau)$ as: 
\begin{eqnarray}\label{Vj}
U_j(\Delta \tau)=    e^{-i \frac{H_0}{D_{12}} \Delta\tau^{(1)}}  e^{-i \frac{H_0+\Omega_j}{D_{12}} \Delta\tau^{(2)}} \approx
  e^{-i \frac{H_0}{D_{12}} \Delta\tau^{(1)}}e^{-i \frac{\Omega_j}{D_{12}} \Delta\tau^{(2)}}.  
\end{eqnarray}
We denote the evolution operator by $U^{(0)}$ to distinguish it from the Trotterized operator $U^{(n)}$, $n>0$. 
We also introduce the parameter $\varepsilon$,
\begin{eqnarray}\label{var0}
\varepsilon=\frac{\Delta\tau^{(2)}}{\Delta\tau},
\end{eqnarray}
that characterizes the relative duration of $\Delta \tau^{(2)}$.

Thus, we have
\begin{eqnarray}\label{Vj2}
U^{(0)}(\tau_{\mathrm{reg}},\varepsilon) = \prod_{j=1}^{K_\omega} U_{j}\left(\frac{\tau_{\mathrm{reg}}}{K_\omega}
,\varepsilon\right) .
\end{eqnarray}
After replacing $U\to U^{(0)}$ and including the parameter $\varepsilon$ into the list of the arguments, expressions for scale factors~(\ref{K})  become
\begin{eqnarray}\label{lambda3}
&& \lambda^{(0)}_{N_RM_R}(\omega,\tau_{\mathrm{reg}},\varepsilon)= 
\lambda^{(0)}_{N_R0_R}(\omega,\tau_{\mathrm{reg}},\varepsilon)(\lambda_{M_R0_R}^{(0)}(\omega,\tau_{\mathrm{reg}},\varepsilon))^*,\;\;N_R\neq M_R,
\\\nonumber
&& \lambda^{(0)}_{N_R0_R}(\omega,\tau_{\mathrm{reg}},\varepsilon) = U^{(0)}_{0_S0_{TL} N_R; N_R;0_{TL} 0_R}(\omega,\tau_{\mathrm{reg}},\varepsilon),
\end{eqnarray}
and Eq.~(\ref{P1}) takes the form 
\begin{eqnarray}\label{constr14}
&&
U^{(0)}_{0_S 0_{TL}N_R; I_S0_{TL}0_R}(\omega,\tau_{\mathrm{reg}},\varepsilon)  =0,\;\; N_R\neq I_S ,\\\nonumber
&&\sum_{N_S,N_{TL}} U^{(0)}_{N_SN_{TL}N_R; I_S0_{TL}0_R}(\omega,\tau_{\mathrm{reg}},\varepsilon) (U^{(0)}(\omega,\tau_{\mathrm{reg}},\varepsilon))^\dagger_{J_S 0_{TL}0_R; N_SN_{TL}M_R} =0,\\\nonumber
&&N_R \neq I_S,\;\; M_R\neq J_S,\;\; N_R\neq M_R,
\end{eqnarray}
We denote different solution $\omega$ of system (\ref{constr14}) by $\omega^{(0;m)}$, $m=1,2,\dots$.

Thus, Model 2 reduces the real Hamiltonian (\ref{HHj}) to the simplified version~(\ref{HHj2}). Now, similar to Model 1, we have to estimate the accuracy of such simplification.
For this, we introduce the characteristics $S^{(0)}_i$, $i=1,\dots,4$ (similar to  $S^{(n)}_i$, $n>0$, in Sec.~\ref{Section:Trot}) as functions of $\tau_{\mathrm{reg}}$ and $\varepsilon$:
\begin{eqnarray}\label{2epsilon}
&&S^{(0)}_1(\tau_{\mathrm{reg}},\varepsilon) =\min_m \max_P P(\omega^{(0;m)},\tau_{\mathrm{reg}},\varepsilon),
\\\label{2epsilon2}&&S^{(0)}_2(\tau_{\mathrm{reg}},\varepsilon) =\min_m\max_{N_R} \left|\lambda^{(0)}_{N_R0_R}(\omega^{(0,m)},\tau_{\mathrm{reg}},\varepsilon)) -
\lambda_{N_R0_R}(\omega^{(0,m)},\tau_{\mathrm{reg}},\varepsilon))\right|,\\\label{2S3}
&&
S^{(0)}_3(\tau_{\mathrm{reg}},\varepsilon) = \max_{0\le\tilde \tau\le \tau_{\mathrm{reg}}} S^{(0)}_1(\tilde \tau,\varepsilon),
\\\label{2S4}
&&
S^{(0)}_4(\tau_{\mathrm{reg}},\varepsilon) = \max_{0\le\tilde \tau\le \tau_{\mathrm{reg}}} S^{(0)}_2(\tilde \tau,\varepsilon).  
\end{eqnarray}
Finally,  we introduce the function $S^{(0)}_5$,
\begin{eqnarray}\label{2S5}
S^{(0)}_5(\tau_{\mathrm{reg}},N,\varepsilon) = \max_{0\le\tau\le \tau_{\mathrm{reg}}} \max_{m} \min_{N_R} \left| \lambda^{(0)}_{N_R0_R} (\omega^{(0;m)},\tilde\tau,\varepsilon)\right|,
\end{eqnarray}
which, similar to $S^{(n)}_5$ in Model 1,  shows  how accurately we have restored the transferred state. 
 For  Model 2,  we also introduce the parameters  $\lambda^{(0)}$ and $\tau_0(N)$,
 \begin{eqnarray}\label{lam02}
\lambda^{(0)}= S^{(0)}_5(T,N,\varepsilon) = S^{(0)}_5(\tau_0(N) ,N,\varepsilon), \;\;\;0<\tau_0(N)< T 
 \end{eqnarray}
(with some fixed large enough time-interval $T$ depending on $N$), demonstrating effectiveness of  applying the state-restoring protocol to the chains of different lengths over the $\tau$-interval $T$.

\section{(0,1)-excitation subspace}
\label{Section:examples}
We consider the restricted state space including only the 0- and 1-excitations (so-called (0,1)-excitation state subspace). Then a set of simplifications can be used  in the above formulae. An important feature of this protocol is that we can also restore the diagonal elements of the receiver's density matrix except the diagonal element $\rho^{(R)}_{0_R;0_R}$ corresponding to the probability of the ground state \cite{TZarxiv}. Let us provide some details for this case. 

For any nonzero multi-index  $J$, $|J|=1$.
The $\lambda$-parameters are again defined by Eqs~(\ref{K}), 
but the second equation in (\ref{P1}) now reads
\begin{eqnarray}\label{TNMIJ}
T_{N_RM_R;I_SJ_S}
 &=& U_{0_S0_{TL}N_R; I_S0_{TL}0_R} U^\dagger_{J_S 0_{TL}0_R; 0_S0_{TL}M_R} =0,   \\\nonumber
 &&\;\;N_R\neq I_S,\;\;M_R\neq J_S, \;\;N_R\neq M_R.
\end{eqnarray}
which is equivalent to the first equation in (\ref{P1}).
Therefore,  system (\ref{P1}) reduces to the single  equation:
\begin{eqnarray}\label{constr_red}
 U_{0_S0_{TL}N_R; I_S0_{TL}0_R} =0,\;\;|I_S| =|N_R| =1,\;\; N_R\neq I_S.
\end{eqnarray}
In additoin, for the diagonal elements of  $\rho^{(R)}$, we have representation (\ref{TNMIJ}):
\begin{eqnarray}\label{TNNIJ}
T_{N_RN_R;I_SJ_S}
 &=& U_{0_S0_{TL}N_R; I_S0_{TL}0_R} U^\dagger_{J_S 0_{TL}0_R; 0_S0_{TL}N_R} ,        \;\;N_R\neq 0_R,
\end{eqnarray}
so that
\begin{eqnarray}\label{TNNIJ1}
    T_{N_RN_R;I_SJ_S} =0 \;\;{\mbox{if}} \;\; N_R\neq I_S \;\;{\mbox{or}}\;\; N_R\neq J_S,\\\label{TNNIJ2}
 T_{N_RN_R;N_RN_R} = |\lambda_{N_R 0_R}|^2.
\end{eqnarray}
In addition,
\begin{eqnarray}
T^{(n)}_{0_R0_R;I_SJ_S} &=& \sum_{{{N_S,N_{TL}}\atop{|N_S|+|N_{TL}|=}}\atop{|I_S|=|J_S|}} U_{N_SN_{TL}0_R; I_S0_{TL}0_R} (U^{(n)})^\dagger_{J_S 0_{TL}0_R; N_SN_{TL}0_R}+\delta_{I_S0_S}\delta_{J_S0_S}.
\end{eqnarray}
Thus, according to (\ref{rhoR}), the diagonal elements  of $\rho^{(R)}$ are all restored except for the element $\rho^{(R)}_{0_R0_R}$.

Formulae defining $S_i^{(n)}$, $n>0$, $i=2,3,4,5$,  (\ref{epsilon2}), (\ref{S3}),  (\ref{S4}) and (\ref{S5}) as well as formulae
defining $S_i^{(0)}$, $i=2,3,4,5$ (\ref{2epsilon2}), (\ref{2S3}), (\ref{2S4})  and (\ref{2S5}) remain the same, while (\ref{epsilon})and (\ref{2epsilon})  become simpler:
\begin{eqnarray}\label{S1}
&&
S^{(n)}_1(\tau_{\mathrm{reg}}) =\min_m \max_{N_R\neq I_S} U_{0_S0_{TL}N_R; I_S0_{TL}0_R}(\omega^{(n;m)},\tau_{\mathrm{reg}}),\;\;n>0,\\\label{2S1}
&&
S^{(0)}_1(\tau_{\mathrm{reg}},\varepsilon) =\min_m \max_{N_R\neq I_S} U_{0_S0_{TL}N_R; I_S0_{TL}0_R}(\omega^{(0;m)},\tau_{\mathrm{reg}},\varepsilon).
\end{eqnarray}

\subsection{Numerical simulations}
\label{Section:Num}
In all examples of this section we deal with (0,1)-excitation evolution. 
We set $N^{(S)} = N^{(R)} = N^{(ER)}$ and consider a homogeneous chain with sender
and receiver of  2 or 3 (Sec.~\ref{Section:NnS3}) spins. To simplify the control of boundness of $|a_{ij}|$ obtained as  numerical 
solutions of system (\ref{constr_red}), we represent $a_{ij}$ as
\begin{eqnarray}\label{ta}
a_{ij} =2 \sin \, \tilde a_{ij}.
\end{eqnarray}
Thus $-2\le a_{ij}\le 2$.

First we give a detailed study of the short chain $N=6$ and show that both  approximating models, Model 1 and Model 2,
are quite reasonable at, respectively, large enough Trotterization number $n$ and small parameter $\varepsilon$. 

\subsubsection{Two-qubit sender $N^{(S)}=N^{(R)}=N^{(ER)}=2$ in spin chain of $N=6$ nodes.}
We consider the short chain of $N=6$ spins with $N^{(S)}=N^{(R)}=N^{(TL)}=2$.
Since $N^{(S)}=2$, there are  two nonzero Larmor  frequencies associated with two last nodes of the chain: 
$\omega_{6}$ and $\omega_{5}$. Eq.(\ref{omega2}) yields $\omega_6=-\omega_5$, thus we have only one free Larmor frequency  $\omega_5$ which will be used as a control tool. The system (\ref{constr_red}) consists of two complex equations 
\begin{eqnarray}\label{constr_red_6}
&&
U_{\{0,0\}\{0,0\}\{0,1\};\{1,0\}\{0,0\}\{0,0\}}(\omega)= 0,\\\nonumber
&&
U_{\{0,0\}\{0,0\}\{1,0\};\{0,1\}\{0,0\}\{0,0\}}(\omega)= 0,
\end{eqnarray}
and there are only two independent scale factors
\begin{eqnarray}\label{lam_num}
\lambda_{\{1,0\}\{0,0\}} = U_{\{0,0\}\{0,0\}\{1,0\};\{1,0\}\{0,0\}\{0,0\}}(\omega),\\\nonumber
\lambda_{\{0,1\}\{0,0\}} = U_{\{0,0\}\{0,0\}\{0,1\};\{0,1\}\{0,0\}\{0,0\}}(\omega).
\end{eqnarray}

To solve (\ref{constr_red_6}), we need at least four parameters $a_{5j}$, $j=1,\dots,4$, generated by the step-wise $\tau$-dependence of $\omega_5$ as given in (\ref{LF}). Numerical simulations show that four parameters $a_{5j}$ are enough, so that we  fix $K_\omega=4$ and
$w=\{a_{51},\dots,a_{54}\}$ in (\ref{constr_red_6}) and (\ref{lam_num}).
In addition, since  all the time intervals $\Delta\tau=\tau_j-\tau_{j-1}$ in (\ref{LF}) are of equal duration, then $\Delta\tau= {\tau_{\mathrm{reg}}/4}$.
For optimization of $S_i^{(n)}$, $i=1,\dots,4$, 
we take 1000 different solutions of system (\ref{constr_red_6}) $a^{(m)}_{5j}$ or $\tilde a^{(m)}_{5j}$ (according to representation (\ref{ta})), $j=1,\dots,4$, $m=1,\dots,1000$.  

\paragraph{Model 1.}
\label{Section:Model1}
We approximate the evolution operator with the Trotterization formula (\ref{Vtr}) setting the  Trotterization number $n=10$, 20, 30, 60. 
Parameters $\omega^{(n;m)}$, $m=1,\dots,1000$, are  solutions of the system 
\begin{eqnarray}\label{constr_red_6_Tr}
&&
U^{(n)}_{\{0,0\}\{0,0\}\{0,1\};\{1,0\}\{0,0\}\{0,0\}}(\omega^{(n;m)})= 0,\\\nonumber
&&
U^{(n)}_{\{0,0\}\{0,0\}\{1,0\};\{0,1\}\{0,0\}\{0,0\}}(\omega^{(n;m)})= 0,
\end{eqnarray}
which is obtained from (\ref{constr_red_6}) by replacing $U \to U^{(n)}$, $\omega \to \omega^{(n;m)}$.
Formulae (\ref{S1}) and (\ref{epsilon2}) for $S_1^{(n)}$ and $S_2^{(n)}$ read, respectively, 
\begin{eqnarray}\label{Tr_S1}
S_1^{(n)}(\tau_{\mathrm{reg}}) &=& \min_m \max\Big(\left|U_{\{0,0\}\{0,0\}\{0,1\};\{1,0\}\{0,0\}\{0,0\}}(\omega^{(n;m)},\tau_{\mathrm{reg}})\right|, \\\nonumber
&&\hspace{1.9cm}
\left|U_{\{0,0\}\{0,0\}\{1,0\};\{0,1\}\{0,0\}\{0,0\}}(\omega^{(n;m)},\tau_{\mathrm{reg}})\right|\Big),\\\label{Tr_S2}
S_2^{(n)}(\tau_{\mathrm{reg}}) &=&
\min_m \max\Big(\left|\lambda^{(n)}_{\{1,0\}\{0,0\}}(\omega^{(n;m)},\tau_{\mathrm{reg}}) - \lambda_{\{1,0\}\{0,0\}}(\omega^{(n;m)},\tau_{\mathrm{reg}})\right|,\\\nonumber
&&\hspace{1.9cm}
\left|\lambda^{(n)}_{\{0,1\}\{0,0\}}(\omega^{(n;m)},\tau_{\mathrm{reg}}) - \lambda_{\{0,1\}\{0,0\}}(\omega^{(n;m)},\tau_{\mathrm{reg}})\right|\Big).
\end{eqnarray}
where
\begin{eqnarray}\label{lam_num2}
\lambda^{(n)}_{\{1,0\}\{0,0\}} = U^{(n)}_{\{0,0\}\{0,0\}\{1,0\};\{1,0\}\{0,0\}\{0,0\}}(\omega^{(n;m)}),\\\nonumber
\lambda^{(n)}_{\{0,1\}\{0,0\}} = U^{(n)}_{\{0,0\}\{0,0\}\{0,1\};\{0,1\}\{0,0\}\{0,0\}}(\omega^{(n;m)}).
\end{eqnarray}
Formulae (\ref{S3}) and (\ref{S4}) for $S_3^{(n)}$ and  $S_4^{(n)}$ remain the same.
Formula (\ref{S5}) for $S^{(n)}_5$ and (\ref{lamn2}) for $\lambda^{(n)}$ take the forms
\begin{eqnarray}\label{Tr_S5}
&&
S^{(n)}_5(\tau_{\mathrm{reg}},N)=\max_{0\le\tilde\tau\le\tau_{\mathrm{reg}}}\max_m \min\Big(\left|\lambda^{(n)}_{\{1,0\}\{0,0\}}(\omega^{(n;m)},\tilde \tau)\right|,
\left|\lambda^{(n)}_{\{0,1\}\{0,0\}}(\omega^{(n;m)},\tilde \tau)\right|\Big),\\\label{lamn3}
&&
\lambda^{(n)} = S^{(n)}_5(\tau_0(N),N),\;\; 0\le\tau_0\le 10 N,
\end{eqnarray}
where we take $T=10 N$ in calculating $S^{(n)}_5$.

\begin{figure*}[!]
\centering
    \begin{subfigure}[c]{0.49\textwidth}
    \centering
    \includegraphics[width=\textwidth]{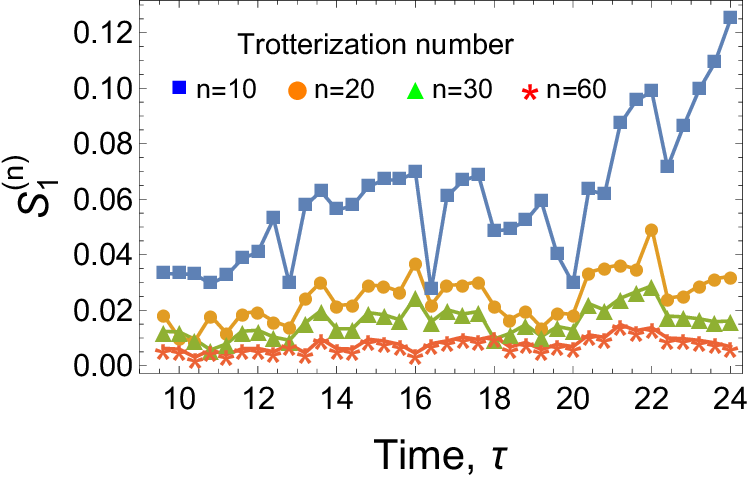}
  \caption{}
  \end{subfigure}
  \begin{subfigure}[c]{0.49\textwidth}
  \centering
  \includegraphics[width=\textwidth]{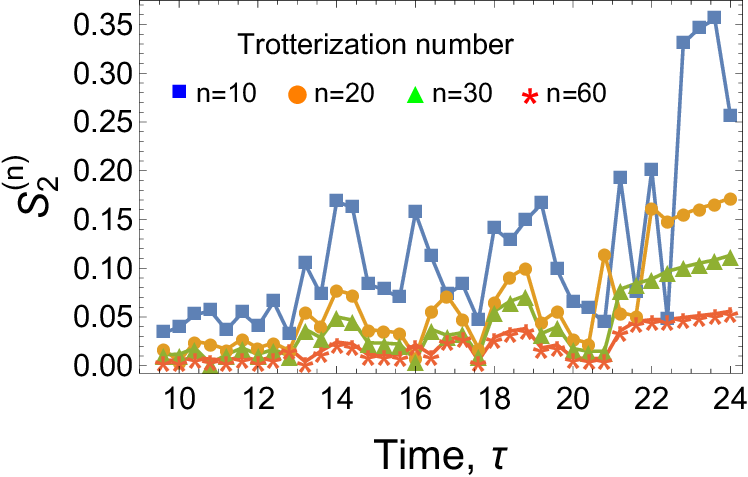}
  \caption{}
  \end{subfigure}
\caption{The $\tau$-dependence of the characteristics $S^{(n)}_1$ (a) and $S^{(n)}_2$ (b). Here $N=6$, $N^{(S)}=N^{(R)}=2$.}  
\label{Fig:S1S2}
\end{figure*}

Characteristics $S^{(n)}_i$, $i=1,2$, are shown in Fig.~\ref{Fig:S1S2}. We see that the Trotterization yields a good approximation for the evolution over relatively short time intervals $\sim 24$ and error of approximation decreases with an increase in Trotterization number. 

\begin{figure*}[!]
\centering
    \begin{subfigure}[c]{0.49\textwidth}
    \centering
    \includegraphics[width=\textwidth]{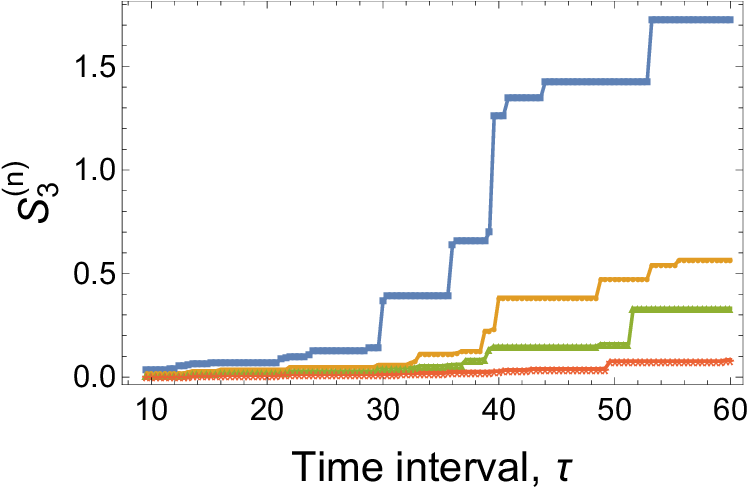}
  \caption{}
  \end{subfigure}
  \begin{subfigure}[c]{0.49\textwidth}
  \centering
  \includegraphics[width=\textwidth]{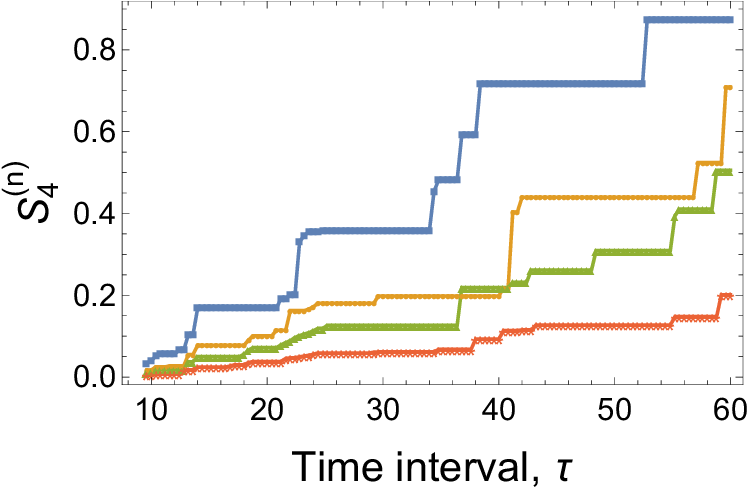}
  \caption{}
  \end{subfigure}
\caption{The $\tau$-dependence of  the characteristics $S^{(n)}_3$ (a) and $S^{(n)}_4$ (b). The Trotterization number $n$ decreases in upward direction taking values 60, 30, 20, 10.
Here $N=6$, $N^{(S)}=N^{(R)}=2$.}  
\label{Fig:S3S4}
\end{figure*}

\begin{figure*}[!]
\centering
\includegraphics[scale=0.65]{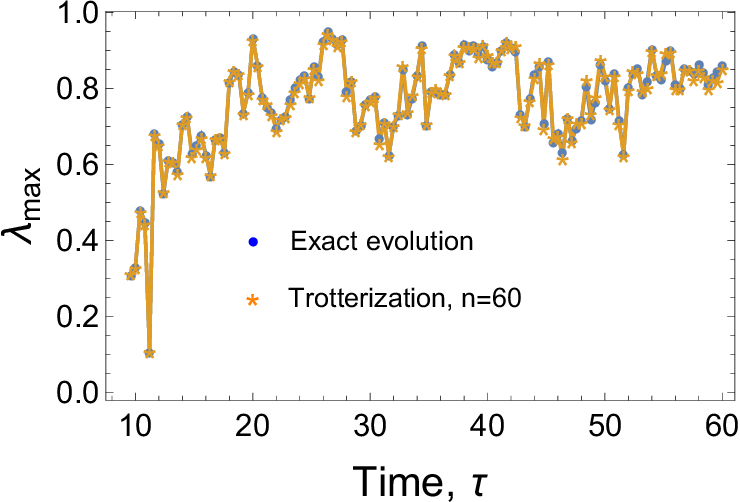}
\caption{The $\tau$-dependence of the maximal scale factor $\lambda_{\max}=\max(\lambda_{01;00},\lambda_{10,00})$. Both the exact and Troterrized ($n=60$) evolutions are shown. Here $N=6$, $N^{(S)}=N^{(R)}=2$.
}  
\label{Fig:max}
\end{figure*}

Approximation over longer time intervals is worse, as shown by   $S^{(n)}_i$, $i=3,4$  in Fig.~\ref{Fig:S3S4}.
Step-wise behavior of $S_3^{(n)}$ and $S_4^{(n)}$ is due to the finite $\tau$-step (which equals 1/10) taken for constructing the characteristics $S^{(n)}_i$, $i=1,\dots,4$. Comparison of the exact evolution of $\lambda_{\max}=\max(\lambda_{01;00},\lambda_{10,00})$ with the approximated  evolution (with Trotterization number $n=60$) is shown in Fig.~\ref{Fig:max} for the optimized parameters $\omega^{(n)}$. We see that the approximation is rather good over the whole considered time interval $0\le{\tau_{\mathrm{reg}}}\le 60$.  

\begin{figure*}[!]
\centering
\includegraphics[scale=0.65]{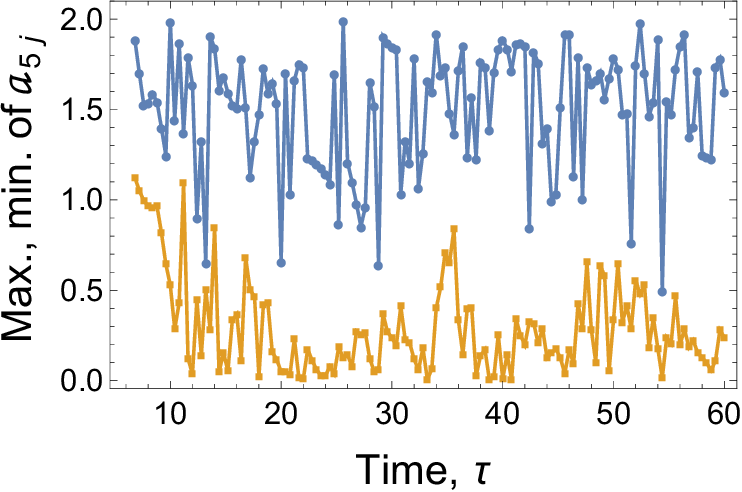}
\caption{The maximal (upper curve) and minimal (lower curve) values of $a_{5j}$, $j=1,\dots,4$ obtained in constructing $S_1^{(60)}$. Here $N=6$, $N^{(S)}=N^{(R)}=2$.
}  
\label{Fig:MM}
\end{figure*}

For completeness, we characterize the values of the parameters $a_{5j}$ used for  constructing $S_1^{(60)}$ (with the Trotterization number $n=60$). We represent the maximal and minimal (by  absolute values) optimized ({minimizing}  $S_1^{(n)}$) parameters $a_{5j}$, $j=1,\dots,4$ 
($\min_j\, |a_{5j}|$, $\max_j\, |a_{5j}|$)  in dependence on the time instant $\tau$ in Fig. \ref{Fig:MM}. By construction, these values are inside of the interval $[0,2]$. 

\paragraph{Model 2.} 
\label{Section:M2}

Formulae for $S^{(0)}_{i}$, $i=1,\dots,5$ and for $\lambda^{(n)}$
are similar to Eqs.~(\ref{Tr_S1}), (\ref{Tr_S2}), (\ref{S3}), (\ref{S4}), (\ref{Tr_S5}) and (\ref{lamn3}) with replacement 
of the superscript $n$ by 0 and adding the parameter
$\tilde \varepsilon$ into the list of the arguments.
In particular,
\begin{eqnarray}\label{Tr_S52}
&&
S^{(0)}_5(\tau_{\mathrm{reg}},N,\tilde \varepsilon)=\max_{0\le\tilde\tau\le\tau_{\mathrm{reg}}}\max_m \min\Big(\left|\lambda^{(0)}_{\{1,0\}\{0,0\}}(\omega^{(n;m)},\tilde \tau)\right|,
\left|\lambda^{(0)}_{\{0,1\}\{0,0\}}(\omega^{(n;m)},\tilde \tau)\right|\Big),\\\label{lamn32}
&&
\lambda^{(0)}= S^{(0)}_5(\tau_0(N),N),\;\; 0\le\tau_0\le 10 N,
\end{eqnarray}

Let us study the dependence of $S_i^{(0)}$, $i=1,\dots,5$, on $\varepsilon$.
We proceed as follows.
First, we set  $ \varepsilon=1/2$, $\Delta\tau^{(2)}= \Delta\tau^{(1)}$, $\Delta\tau=2  \Delta\tau^{(1)}$ and  find parameters $\omega^{(0;m)}=\{a_{51}^{(m)},\dots,a_{54}^{(m)}\} $  as solutions of  system
\begin{eqnarray}\label{constr_red_6_2}
&&
U^{(0)}_{\{0,0\}\{0,0\}\{0,1\};\{1,0\}\{0,0\}\{0,0\}}(\omega^{(0;m)})= 0,\\\nonumber
&&
U^{(0)}_{\{0,0\}\{0,0\}\{1,0\};\{0,1\}\{0,0\}\{0,0\}}(\omega^{(0;m)})= 0,\;\;m=1,\dots 1000,
\end{eqnarray}
obtained from (\ref{constr_red_6}) after replacement $U\to U^{(0)}$.
It is remarkable that having $\Omega_j$ constructed with $\varepsilon =1/2$ ($\Delta\tau^{(2)}= \Delta\tau^{(1)}$) we can deform the obtained result to various $\varepsilon$ with appropriate shrinking of $\tau$ without additional  simulating spin dynamics.  	
For a further analysis, it is convenient to introduce the parameter $\tilde\varepsilon$ instead of $\varepsilon$ given in  (\ref{var0}) and $\tilde\varepsilon$-dependent subinterval $\Delta\tau^{(2)}(\tilde\varepsilon)$ by the formulae
\begin{eqnarray}\label{te}
\Delta\tau^{(2)}(\tilde\varepsilon)  =\Delta\tau_1 \tilde\varepsilon \;\;\Rightarrow \;\;
\tilde \varepsilon = \frac{\Delta\tau^{(2)}(\tilde\varepsilon) }{\Delta\tau_1}.
\end{eqnarray}
Then we have
\begin{eqnarray}
\Omega_j \Delta \tau^{(1)} = \frac{\Omega_j}{\tilde\varepsilon} \Delta\tau^{(2)}(\tilde\varepsilon).
\end{eqnarray}
Therefore, we may use  the replacement  $\displaystyle a_{ij} \to \frac{a_{ij}}{\tilde\varepsilon}$ in the Hamiltonian (\ref{HHj}), so that condition (\ref{avar}) is satisfied and this condition provides  the structure (\ref{HHj2}) for the Hamiltonian.
Because of (\ref{te}), the interval $\Delta\tau$ becomes also $\tilde\varepsilon$-dependent,
\begin{eqnarray}\label{dt12}
\Delta \tau(\tilde\varepsilon) = \Delta\tau^{(1)} +\Delta\tau^{(2)}(\tilde\varepsilon).
\end{eqnarray}
It can be simply verified that
\begin{eqnarray}
\tilde \varepsilon = \frac{\varepsilon}{1-\varepsilon}\stackrel{\varepsilon\ll 1}{ \approx} \varepsilon .
\end{eqnarray}
In this way, we can vary $\tilde \varepsilon$ scaling $\Omega_j$ and thus scaling $a_{ij}$. In studying the functions $S_i^{(0)}$ , $i=1,\dots,5$, below we use $\tilde\varepsilon$ instead of $\varepsilon$.

\begin{figure*}[!]
\centering
\begin{subfigure}[c]{0.49\textwidth}
\centering
\includegraphics[width=\textwidth]{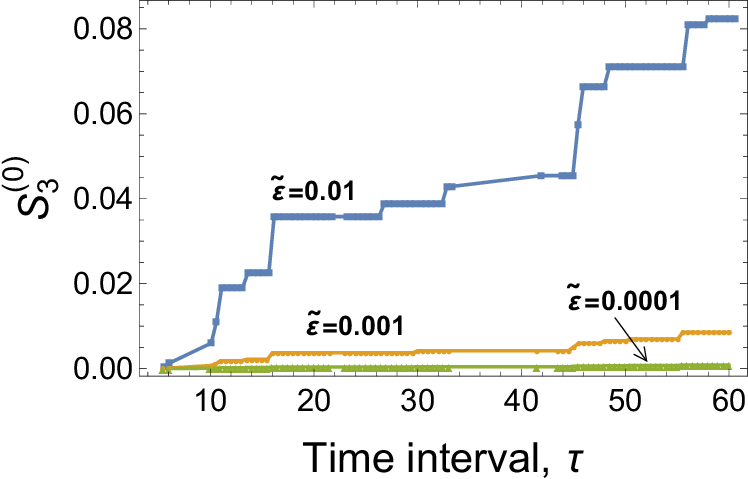}
\caption{}
\end{subfigure}
\begin{subfigure}[c]{0.49\textwidth}
\centering
\includegraphics[width=\textwidth]{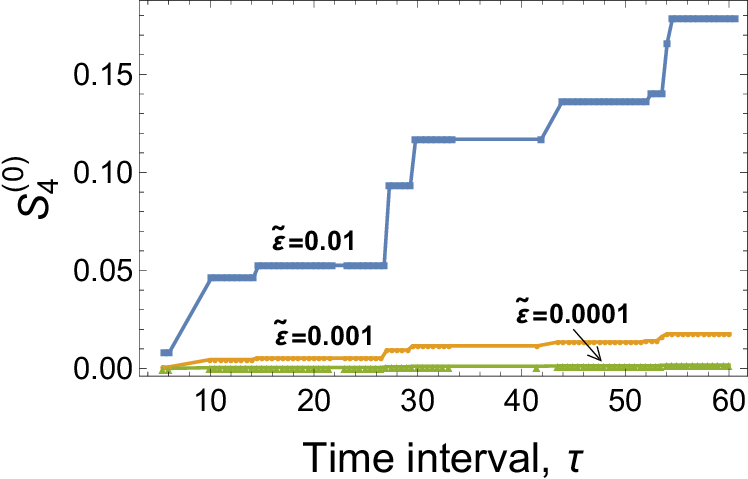}
\caption{}
\end{subfigure}
\caption{The $\tau$-dependence of  the characteristics $S^{(0)}_3$ (a) and $S^{(0)}_4$ (b) for different  $\tilde \varepsilon =0.01,\; 0.001,\; 0.0001$. Here $N=6$, $N^{(S)}=N^{(R)}=2$.
}  
\label{Fig:2S3S4}
\end{figure*}

The graphs of $S^{(0)}_{i}$, $i=1,\dots,4$ are quite similar to the graphs of $S^{(n)}_{i}$, $i=1,\dots,4$ in Sec.~\ref{Section:Model1}, see Fig.~\ref{Fig:S1S2} and Fig.~\ref{Fig:S3S4}. For instance, the graphs of $S^{(0)}_{i}$, $i=3,4$ 
as functions of  $\tau$ for different  $\tilde \varepsilon =0.01,\; 0.001,\; 0.0001$ are shown in Fig.~\ref{Fig:2S3S4}.

Notice that since the parameters  $a_{5j}$ are represented in the form~(\ref{ta}), 
they are  bounded by the condition $|a_{5j}|\in [0,2]$. Of course, the scaled parameters $\displaystyle\frac{a_{5j}}{\tilde\varepsilon}$ do not obey this constraint. 

\subsubsection{Two- and three-qubit state restoring in long chains ($N\le 30$)}
To demonstrate the applicability of the  state-restoring protocol to long communication lines, we consider the $N$-dependence of the  function $S^{(0)}_5$ in the state-restoring of two- and three-qubit states transferred using long communication lines $N\le 30$.
In this section we use only Model 2 and take $\tilde \varepsilon=0.0001$.

\begin{figure*}[!]
\centering
  \includegraphics[width=0.49\textwidth]{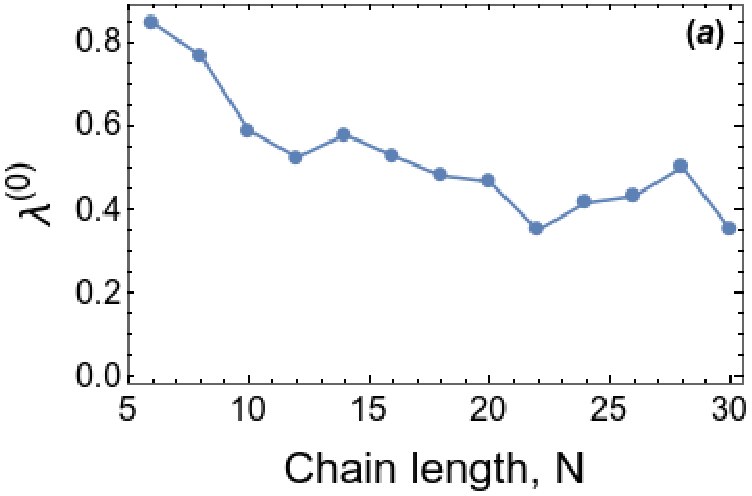}
  \includegraphics[width=0.49\textwidth]{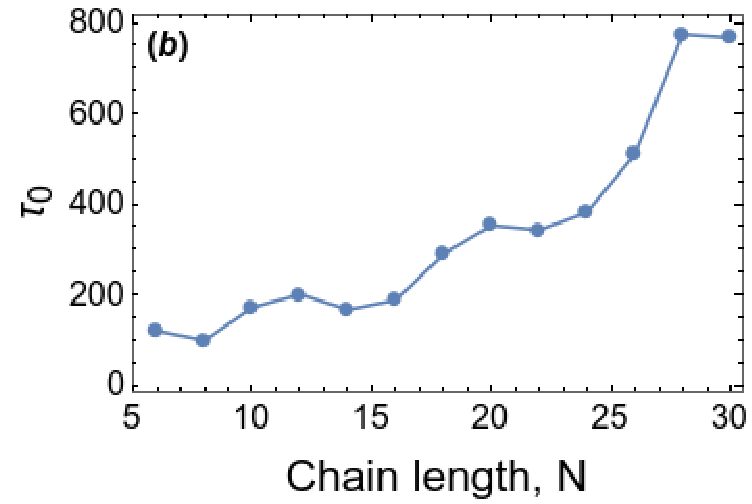}
  \includegraphics[width=0.49\textwidth]{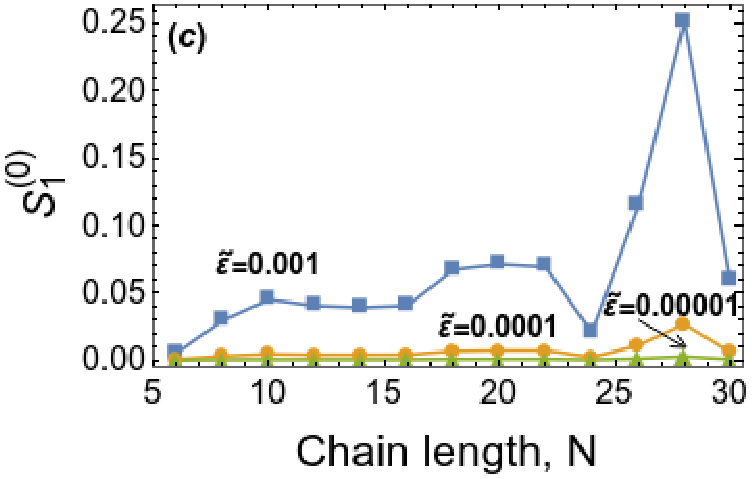}
  \includegraphics[width=0.49\textwidth]{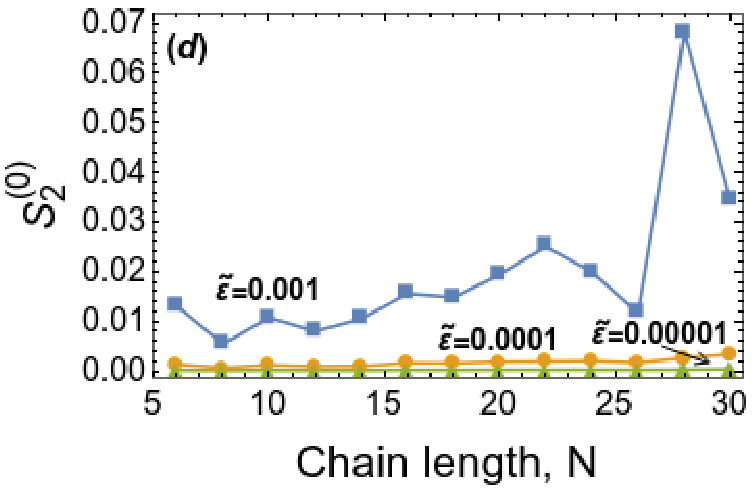}
  \includegraphics[width=0.49\textwidth]{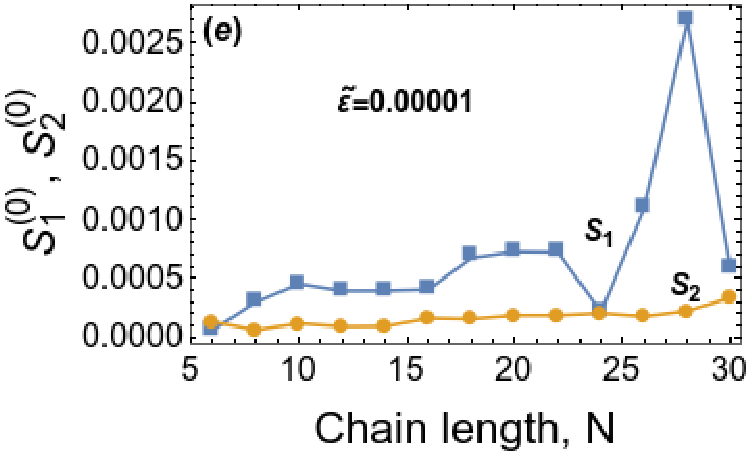}
\caption{$N^{(S)}=N^{(R)}=2$; the scale factor $\lambda^{(0)}(N)=S_5^{(0)}(\tau_0,N)$,
$0\le \tau_0(N) \le 30 N$, (a) and the corresponding  time instant $\tau_0(N)$ (b)  as  functions of the chain length $N$  (even $N$); optimization is over 1000 solutions of Eq.~(\ref{constr_red_6_2}). {Characteristics $S^{(0)}_1$ and $S^{(0)}_2$ at time instant $\tau_0(N)$ for different values of parameter $\tilde\varepsilon$ are shown, respectively,  in (c) and (d), also for even values of $N$. Graphics of $S^{(0)}_1$ and $S^{(0)}_2$ for $\tilde\varepsilon=0.00001$ are shown in (e) }
}  
  \label{Fig:CT}
\end{figure*}

\paragraph{Two-qubit sender, $N^{(S)}=N^{(R)}=N^{(ER)}=2$.} 
We plot  $\lambda^{(0)}$, given in (\ref{lam02})  (fixing $K_\omega=5$ instead of 4 in Sec.~\ref{Section:M2}) as a function of $N$ in Fig.~\ref{Fig:CT}. For optimization  $S_5^{(0)}(T,N)$, we take $T=30 N$ and 1000 different solutions $\omega^{(0;m)}$ of the system~(\ref{constr_red_6_2}).
These graphs demonstrate generic decrease of $\lambda_{\min}(N)$ and increase of $\tau_0(N)$ with $N$. The fact that points of these graphs do not belong to smooth lines is explained, basically, by difficulties in the global optimization  of $S^{(0)}_5$ according to its definition~(\ref{Tr_S52}). {We also compute and plot characteristics $S^{(0)}_1$ and $S^{(0)}_2$ at time instant $\tau_0(N)$ for various values of $\tilde\varepsilon$. Characteristics $S^{(0)}_3$ and $S^{(0)}_4$ are not shown because they are  associated with global $\tau$-evolution of a particular chain with fixed length $N$.}

\paragraph{Three-qubit sender, $N^{(S)}=N^{(R)}=N^{(ER)}=3$.}
\label{Section:NnS3}

Since $N^{(ER)}=3$, there are three nonzero Larmor frequencies, 
$\omega_{7}$, $\omega_{6}$ and $\omega_{5}$. 
Eq.~(\ref{omega2}) yields $\omega_7=-\omega_6-\omega_5$, 
thus we have two free Larmor frequency  $\omega_5$ and $\omega_6$ 
which can be used as a control tool. 
The system (\ref{constr_red}) consists of 6 complex equations, which are given by (for Model 2) 
\begin{eqnarray}\label{constr_red_62}
U^{(0)}_{0_3 0_{TL} \{0,0,1\};\{0,1,0\} 0_{TL} 0_3}(\omega^{(0;m)})= 0, &&
U^{(0)}_{0_3 0_{TL} \{0,0,1\};\{1,0,0\} 0_{TL} 0_3}(\omega^{(0;m)})= 0,\\\nonumber
U^{(0)}_{0_3 0_{TL} \{0,1,0\};\{0,0,1\} 0_{TL} 0_3}(\omega^{(0;m)})= 0, &&
U^{(0)}_{0_3 0_{TL} \{0,1,0\};\{1,0,0\} 0_{TL} 0_3}(\omega^{(0;m)})= 0,\\\nonumber
U^{(0)}_{0_3 0_{TL} \{1,0,0\};\{0,0,1\} 0_{TL} 0_3}(\omega^{(0;m)})= 0,&&
U^{(0)}_{0_3 0_{TL} \{1,0,0\};\{0,1,0\} 0_{TL} 0_3}(\omega^{(0;m)})= 0,
\end{eqnarray}
where $0_3=\{0,0,0\}$ and $\omega^{(0;m)}$, $m=1,2,\dots$,  are different solutions  of system (\ref{constr_red_62}).
We have 3 independent scale parameters
\begin{eqnarray}\label{lam_num22}
\lambda^{(0)}_{\{1,0,0\} 0_3} = U^{(0)}_{0_3 0_{TL} \{0,0,1\};\{0,0,1\} 0_{TL} 0_3}(\omega^{(0;m)}),\\\nonumber
\lambda^{(0)}_{\{0,1,0\} 0_3} = U^{(0)}_{0_3 0_{TL} \{0,1,0\};\{0,1,0\} 0_{TL} 0_3}(\omega^{(0;m)}),\\\nonumber
\lambda^{(0)}_{\{0,1,0\} 0_3} = U^{(0)}_{0_3 0_{TL} \{1,0,0\};\{1,0,0\} 0_{TL} 0_3}(\omega^{(0;m)}).
\end{eqnarray}

To satisfy system (\ref{constr_red_62}), we take $K_\omega=7$, so that there are 14 free parameters  
$a_{5j}$, $a_{6j}$,  $j=1,\dots,7$. Eqs.(\ref{constr_red_62}) are considered as a system of equations for these 14 parameters. 

\begin{figure*}[!]
\centering
\includegraphics[width=0.49\textwidth]{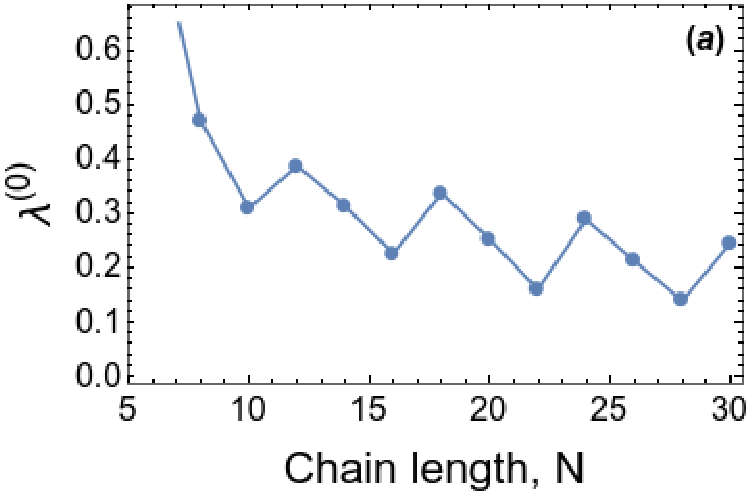}
  \includegraphics[width=0.49\textwidth]{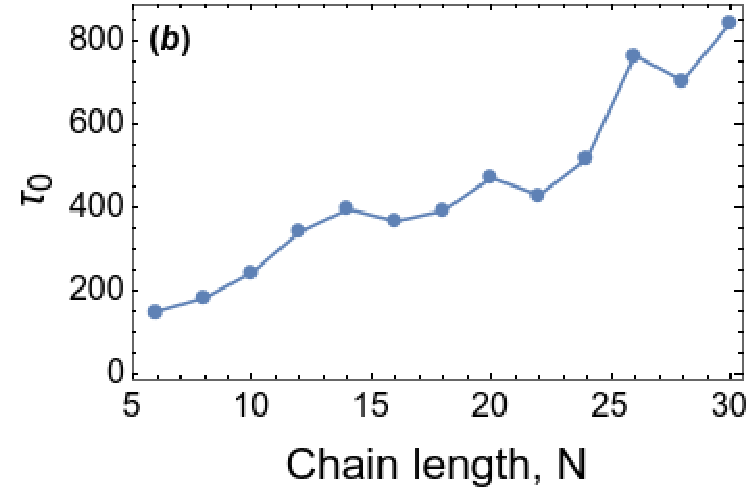}
  \includegraphics[width=0.49\textwidth]{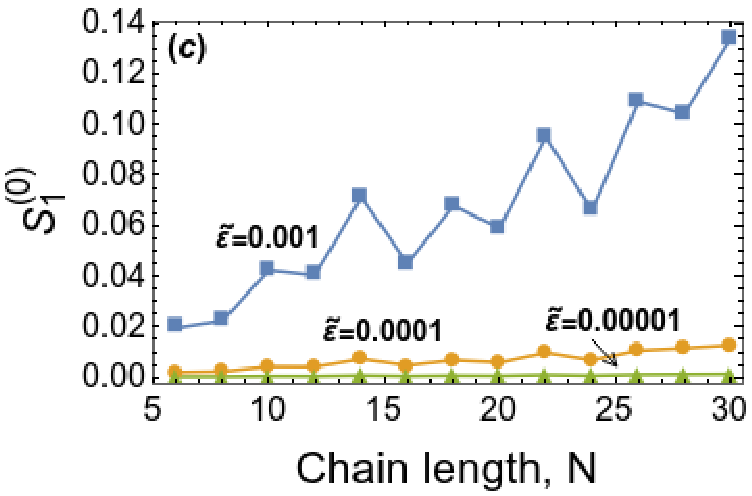}
  \includegraphics[width=0.49\textwidth]{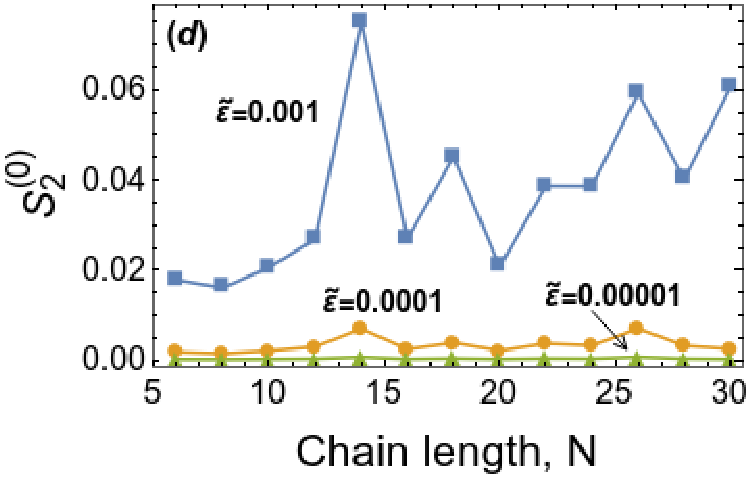}
  \includegraphics[width=0.49\textwidth]{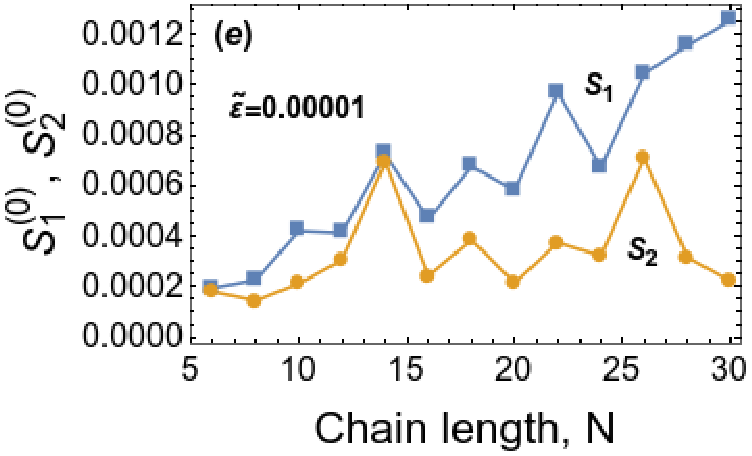}
\caption{$N^{(S)}=N^{(R)}=3$; the scale factor $\lambda^{(0)}(N)=S_5^{(0)}(\tau_0(N),N)$, {$0\le \tau_0(N) \le 30 N$} (a) and corresponding  time instant $\tau_0(N)$ (b) as  functions of chain length $N$ computed for even values of $N$; optimization is over 2000 solutions of Eq.~(\ref{constr_red_62}). 
{Characteristics $S^{(0)}_1$ and $S^{(0)}_2$ at time instant $\tau_0(N)$ computed for various  values of the parameter $\tilde\varepsilon$ are shown, respectively, in (c) and (d), also computed for even values of $N$. Values of $S^{(0)}_1$ and $S^{(0)}_2$ for $\tilde\varepsilon=10^{-5}$ are also shown (in addition to subfigures (c) and (d)) separately in subfigure~(e).}}  
\label{Fig:CT2}
\end{figure*}

Of course, solution of this system is not unique.
For optimization in formulas for $S_5^{(0)}(T,N)$, 
we take $T=30 N$ and 2000 different solutions of system (\ref{constr_red_62}) $a^{(m)}_{5j}$, $a^{(m)}_{6j}$ or $\tilde a^{(m)}_{5j}$,
$\tilde a^{(m)}_{6j}$ (according to representation (\ref{ta})), $j=1,\dots,7$, $m=1,\dots,2000$.  
Finally, we plot  $\lambda^{(0)}=S_5^{(0)}(\tau_0(N),N)$ as a function of $N$ in Fig.\ref{Fig:CT2}.
We see that this graph is similar to the graph in Fig.\ref{Fig:CT} for the case $N^{(S)}=2$, so that we do not observe significant difference in the $N$-dependence of the parameters $\lambda^{(0)}$ and $\tau_0$ for the cases of two- and three-node sender/receiver. {Characteristics $S^{(0)}_1$ and $S^{(0)}_2$ computed for various values of the parameter $\tilde\varepsilon$ are shown, respectively, in Fig.~\ref{Fig:CT2}(c) and Fig.~\ref{Fig:CT2}(d).}

\section{Conclusions}
\label{Section:conclusions}
The remote state restoring is a variant of the state transfer. In this process, density matrix of the restored  receiver's state  at some time instant becomes structurally equivalent to the initial sender's density matrix (i.e., the corresponding  elements of these  two density matrices  are proportional to each other) up to (some) diagonal elements. In our paper, for restoring protocol we use an inhomogeneous time-dependent external magnetic field. In the considered case of (0,1)-excitation state transfer, only one diagonal elements of the density matrix can not be restored. This element corresponds to the 0-excitation state subspace.

For approximation of the many-spin dynamics, we use either Trotteri-Suzuki approach or approximation of strong magnetic pulses. We do not use the traditional fidelity as a measure of quality of the state transfer protocol. In the case of state restoring, it is more natural to take the minimal  absolute value of the scale factors ($\lambda$-factors, see Eq.~(\ref{rest}))  as  a characteristics of effectiveness of the protocol. This is reflected in the parameter $S_5^{(n)}$ and associated parameter $\lambda^{(n)}$, where $n>0$ for Model 1 and $n=0$ for Model 2. In addition, applying the approximation Models 1 and 2 we introduce the set of characteristics $S^{(n)}_i$, $i=1,\dots,4$ to estimate the approximation accuracy of these models. Of special interest is Model 2 (Sec.~\ref{Section:Model2}) because it might be realizable via the tool of multiple quantum NMR \cite{Baum}.

{\bf Acknowledgments.} This work was funded by the Ministry of Science and Higher Education (grant number 075-15-2020-788).

{\bf Data availability statement.} All data that support the findings of this study are included within the article (and any supplementary files).


\begin{thebibliography}{99}

\section*{References}

\bibitem{Schleich2016} Schleich W P, Ranade K S, Anton C, Arndt M, Aspelmeyer M, Bayer M, Berg G, Calarco T, Fuchs H, Giacobino E, Grassl M, Hänggi P, Heckl W M, Hertel I-V, Huelga S, Jelezko F, Keimer B, Kotthaus J P, Leuchs G, Lütkenhaus N, Maurer U, Pfau T, Plenio M B, Rasel E M, Renn O, Silberhorn C, Schiedmayer J, Schmitt-Landsiedel D, Sch\"onhammer K, Ustinov A, Walther P, Weinfurter H, Welzl E, Wiesendanger R, Wolf S, Zeilinger A and Zoller P  
2016  {\it Appl. Phys.} B  {\bf 122} (5) 130

\bibitem{Acin2018} Acín A, Bloch I, Buhrman H, Calarco T, Eichler C, Eisert J, Esteve D, Gisin N, Glaser S J, Jelezko F, Kuhr S, Lewenstein M, Riedel M F, Schmidt P O, Thew R, Wallraff A, Walmsley I and Wilhelm F K 
2018 {\it New J. Phys.} , {\bf 20}(8) 080201


\bibitem{BBCJPW}
Bennett C H,  Brassard G,  Cr\'epeau C,  Jozsa R, Peres A and  Wootters W K
1993 {\it Phys. Rev. Lett.} {\bf 70} 1895

\bibitem{BPMEWZ}
 Bouwmeester D,  Pan J-W, Mattle  K, Eibl M,  Weinfurter H, Zeilinger A 
1997 {\it Nature} {\bf 390}  575.

\bibitem{BBMHP}
Boschi D, Branca S, De Martini F,  Hardy L and Popescu S
 1998 {\it Phys. Rev. Lett.} {\bf 80} 1121


\bibitem{Bose}
Bose S  
2003 {\it Phys. Rev. Lett.}
 {\bf 91}   207901

 \bibitem{CDEL}
 Christandl M, Datta N, Ekert  A and Landahl A J
2004 {\it Phys.Rev.Lett.} {\bf 92}  187902

\bibitem{KS}
Karbach P and Stolze J 
2005 {\it Phys.Rev.} A  {\bf 72} 030301(R)

\bibitem{GKMT}
Gualdi  G, Kostak V, Marzoli I and Tombesi P  
2008 {\it Phys.Rev.} A {\bf 78}  022325


\bibitem{GMT}
Gualdi G, Marzoli I, Tombesi P 
2009 {\it New J. Phys.}  {\bf 11}  063038

\bibitem{ZASO}
Zwick A, \'Alvarez G A, Stolze J, Osenda O
2011 {\it Phys. Rev.} A {\bf 84}  022311

\bibitem{ZASO2}
Zwick A,  \'Alvarez G A,
 Stolze J and Osenda O,
2012 {\it Phys. Rev.} A  {\bf 85}  012318

\bibitem{ZenchukJPA2012} 
Zenchuk A I 
2012 {\it J. Phys. A: Math. Theor.}  {\bf 45}(11)
115306


\bibitem{PBGWK} 
Peters N A,  Barreiro J T,  Goggin M E,  Wei T-C and Kwiat P G 
2005 {\it Phys. Rev. Lett.} {\bf 94}  150502

\bibitem{PBGWK2}
Peters N A, Barreiro J T, Goggin M E,  Wei T-C and Kwiat P G
2005 Remote state preparation: arbitrary remote control of photon polarizations for quantum communication 
 (Quantum Communications and Quantum Imaging III, in: Proc. of SPIE, vol. 5893) eds Meyers R E,  Shih Ya, (SPIE, Bellingham, WA)

\bibitem{DLMRKBPVZBW} Dakic  B, Lipp Ya O, Ma X, Ringbauer M, Kropatschek S,  Barz S,  Paterek T, Vedral V,  Zeilinger A, Brukner C and Walther P 
2012 {\bf Nat. Phys.} {\bf 8} 666


\bibitem{PSB} 
 Pouyandeh S, Shahbazi F and Bayat A
2014 {\bf Phys. Rev.} A {\bf 90}
 012337

\bibitem{LH}
Liu L L and  Hwang T 
2014 {\it Quantum Inf. Process.} {\bf 13} 1639

\bibitem{Z_2014}
 Zenchuk A I
2014 {\bf Phys. Rev.} A {\bf 90}  052302

\bibitem{BZ_2015}
Bochkin G A and  Zenchuk A I
2015 {\it Phys. Rev.} A {\bf 91}  062326

\bibitem{Koch2022} 
Koch C P, Boscain U, Calarco T, Dirr G, Filipp S, Glaser S J, Kosloff R, Montangero S, Schulte-Herbrüggen T, Sugny D, Wilhelm F K 
 2022 {\it EPJ Quantum Technol.} {\bf 9}(1) 19

\bibitem{Osborne2004}{
Osborne T J and Linden N 2004 
{\it Phys. Rev.} A {\bf 69} 052315}


\bibitem{SchirmerPRA2009} Schirmer S G and Pemberton-Ross P J 
 2009 {\it Phys. Rev.} A {\bf 80}(3) 030301 

\bibitem{Ashhab2015} {Ashhab S 2015 
{\it Phys. Rev.} A {\bf 92} 062305}

\bibitem{MurphyPRA2010} Murphy M, Montangero S, Giovannetti V and Calarco T 
2010 {\it Phys. Rev.} A  {\bf 82}(2) 022318

\bibitem{Ashhab2012} {Ashhab S, De Groot P C and Nori F 2012 
{\it Phys. Rev.} A {\bf 85} 052327}

\bibitem{OsendaPLA2021} Acosta Coden D S, G\'omez S S, Ferr\'on A and Osenda O 
 2021 {\bf Phys. Lett.} A {\bf 387} 127009 

\bibitem{MorigiPRL2015} Morigi G, Eschner J, Cormick C, Lin Y, Leibfried D, Wineland D J 
2015 {\it Phys. Rev. Lett.} {\bf 115}(20) 200502 

\bibitem{Mohiyaddin2016} Mohiyaddin F A, Kalra R, Laucht A, Rahman R, Klimeck G, Morello A  
2016 {\it Phys. Rev.} B  {\bf 94}(4) 045314 

\bibitem{AubourgJPB2016} Aubourg L and Viennot D 
2016 {\it J. Phys. B: At. Mol. Opt. Phys.}  {\bf 49}(11) 115501

\bibitem{ShanSciRep2018} Shan H J, Dai C M, Shen H Z, Yi X. X. 
2018 Sci Rep  {\bf 8}(1) 13565

\bibitem{PyshkinNJP2018} Pyshkin P V, Sherman E Y, You J Q and Wu L-A  
 2018 {\it New J. Phys.}  {\bf 20}(10) 105006 


\bibitem{FerronPS2022} Ferr\'on A, Serra P and Osenda O. 
2022 {\it Phys. Scr.}  {\bf 97}(11) 115103 

\bibitem{Taminiau2014} Taminiau T H, Cramer J, Van Der Sar T, Dobrovitski V V and Hanson R 
2014 {\it Nature Nanotech.}  {\bf 9}(3) 171

\bibitem{Peng2021} Peng P, Yin C, Huang X, Ramanathan C and  Cappellaro P
2021 {\it Nat. Phys.}  {\bf 17}(4) 444

\bibitem{Uysal2023} Uysal M T, Raha M, Chen S, Phenicie C M, Ourari S, Wang M, Van De Walle C G, Dobrovitski V V and Thompson J D 2023
 {\it PRX Quantum}  {\bf 4}(1) 010323

\bibitem{KuprovBook}
Kuprov I 2023 {\it Spin: From Basic Symmetries to Quantum Optimal Control}  (Springer)


\bibitem{FZ_2017}
Fel'dman E B and Zenchuk A I 
2017 {\it JETP} {\bf 125}(6)  1042

\bibitem{BFZ_Arch2018}
Bochkin G A, Fel'dman E B and  Zenchuk A I
2018 {\it Quant.Inf.Proc.} {\bf 17} 218

\bibitem{Z_2018}
Zenchuk A I  
2018 {\it Phys.Lett.} A {\bf 382}  324

\bibitem{FPZ_2021}
Fel'dman E B,  Pechen A N and Zenchuk A I
2021 {\bf Phys. Let.} A {\bf 413} 127605 

\bibitem{BFLP_2022}
Bochkin G A, Fel'dman E B, Lazarev I D, Pechen A N and
Zenchuk A I  
2022 {\it Quant. Inf. Proc.} {\bf 21}
 261

\bibitem{NCh}
Nielsen M A,  Chuang I L   2010
{\it Quantum Computation and Quantum Information}
(Cambridge:Cambridge Univ. Press.) p~676 

\bibitem{PP_2023}
Petruhanov V and Pechen  A 
2023 {\it J. Phys. A: Math. Theor.} {\bf 56} 305303 
 	
\bibitem{MP_2023} Morzhin  O and Pechen A 
2023 {\it Quant. Inf. Proc.} {\bf 22} 241 

\bibitem{Trotter} Trotter H F  
1959 {\it Proc. Amer. Math. Soc.} {\bf 10}(4) 545

\bibitem{Suzuki} Suzuki M 
1976 {\it Commun. Math. Phys.}  {\bf 51}(2) 183

\bibitem{Baum}
Baum J,  Munowitz M, Garroway A N and Pines A 1985 {\it J. Chem. Phys.}
{\bf 83} 2015 

\bibitem{Khaneja2005} Khaneja N, Reiss T, Kehlet C, Schulte-Herbr\"uggen T and Glaser S J 
 2005 {\it Journal of Magnetic Resonance} {\bf 172}(2) 296

\bibitem{GlaserJPB2011} Schulte-Herbrüggen T, Sp\"orl A, Khaneja N, Glaser S J  
2011 {\it J. Phys. B: At. Mol. Opt. Phys.}  {\bf 44}(15) 154013

\bibitem{Fouquieres2011}
De Fouquieres P, Schirmer S G, Glaser S J, Kuprov I
2011 {\it J.  Magn. Res.} {\bf 212}(2) 412


\bibitem{PechenTannor2011} Pechen A N and Tannor D J
2012 {\it Israel Journal of Chemistry} {\bf 52} 467

\bibitem{Lucarelli2018} Lucarelli D 
2018 {\it Phys. Rev.} A  {\bf 97}(6) 062346

\bibitem{VMP_21} Volkov B O, Morzhin O V and Pechen A N 
2021 {\it J. Phys. A: Math. Theor.} {\bf 54} 215303 

\bibitem{KrausBook} { Kraus K 1983  {\it States, Effects, and Operations: Fundamental Notions of Quantum Theory}  (Springer Berlin Heidelberg)}

\bibitem{Lieb1961} {Lieb E, Schultz T and Mattis D 
 1961 {\it Annals of Physics} {\bf 16}(3) 407}

\bibitem{Porras2004} {Porras D and Cirac J I 
2004 {\it Phys. Rev. Lett.}  {\bf 92}(20) 207901}


\bibitem{Maik2012} {Maik M, Hauke P, Dutta O, Zakrzewski J and Lewenstein M 
2012 {\it New J. Phys.}  {\bf 14}(11) 113006}


\bibitem{PM}
Pereira D and Mueller E J 
2022 {\it Phys. Rev.} A  {\bf 106} 043306

\bibitem{IBRR}
Igl\'oi F,  Blass B, Ro\'osz G. and Rieger H
2018 {\it Phys. Rev.} B {\bf 98} 184415 


\bibitem{TZarxiv}
Tashkeev N A and Zenchuk A I 2023
Remote restoring of (0,1)-excitation states and
concurrence scaling arXiv:2310.04526 [quant-ph]

\end{thebibliography}
\end{document}